\documentclass[a4paper,11pt]{article}
\pdfoutput=1 
\usepackage{aas_macros}
\usepackage{jcappub} 
\usepackage[T1]{fontenc} 
\usepackage{xspace}
\usepackage{xstring}
\usepackage{orcidlink}
\usepackage{multirow}
\usepackage{amsmath}
\usepackage{bbm}
\usepackage{stmaryrd}
\usepackage{bm}
\usepackage{graphicx}  
\usepackage{tikz}      
\usetikzlibrary{fit, positioning, decorations.pathreplacing, shapes.geometric, arrows.meta, calc, 3d, arrows} 
\usepackage{subcaption}
\usepackage{adjustbox}
\usepackage{comment}
\usepackage{color,hyperref}
\usepackage[T1]{fontenc} 
\usepackage{lineno}

\newcommand{\angstrom}{\ensuremath{\mathring{\mathrm{A}}}\xspace}
\newcommand{\lya}{\ensuremath{\text{Ly}\alpha}\xspace}
\newcommand{\lz}{\ensuremath{\ell_z}}

\newcommand{\zlbg}{\texttt{zlbg}\xspace}
\newcommand{\lbgnet}{\texttt{lbgNET}\xspace}
\newcommand{\qnet}{\texttt{QuasarNET}\xspace}
\newcommand{\redrock}{\texttt{redrock}\xspace}

\newcommand{\eline}[2]{%
\ensuremath{[\mathrm{#1}]\,%
\IfSubStr{#2}{,}{\lambda\lambda}{\lambda}%
#2}}

\newcommand{\wl}[1]{%
  \ensuremath{%
    \IfSubStr{#1}{,}{\lambda\lambda}{\lambda}%
    #1%
  }%
}

\makeatletter
\newcommand\ion[2]{%
  #1$\;${%
    \ifx\@currsize\normalsize\small \else
    \ifx\@currsize\small\footnotesize \else
    \ifx\@currsize\footnotesize\scriptsize \else
    \ifx\@currsize\scriptsize\tiny \else
    \ifx\@currsize\large\normalsize \else
    \ifx\@currsize\Large\large
    \fi\fi\fi\fi\fi\fi
    \rmfamily\@Roman{#2}}%
  \relax
}
\makeatother

\renewcommand\LaTeX{%
 \leavevmode
 L%
 \raise.42ex\hbox{%
  \count@=\the\fam
  $\fam\count@\scriptstyle\kern-.3em A$
 }%
 \kern-.15em\TeX
}%

\newcommand{\setzero}{VI-0\xspace} 
\newcommand{\setone}{VI-1\xspace} 
\newcommand{\settwo}{VI-contaminants\xspace}

\newcommand{\ConvBlock}[8]{%
    \coordinate (A) at (#1,#2,#3);
    \coordinate (B) at (#1,#2+#5,#3);
    \coordinate (C) at (#1,#2+#5,#3+#6);
    \coordinate (D) at (#1,#2,#3+#6);
    \coordinate (E) at (#1+#4,#2,#3);
    \coordinate (F) at (#1+#4,#2+#5,#3);
    \coordinate (G) at (#1+#4,#2+#5,#3+#6);
    \coordinate (H) at (#1+#4,#2,#3+#6);
    \fill[#7!40, fill opacity=#8, thick] (A) -- (B) -- (F) -- (E) -- cycle;
    \fill[#7!50, fill opacity=#8] (B) -- (C) -- (G) -- (F) -- cycle;
    \fill[#7!30, fill opacity=#8] (E) -- (F) -- (G) -- (H) -- cycle;
}

\newcommand{\BNBlock}[8]{\ConvBlock{#1}{#2}{#3}{#4}{#5}{#6}{#7}{#8}}
\newcommand{\DenseBlock}[8]{\ConvBlock{#1}{#2}{#3}{#4}{#5}{#6}{#7}{#8}}
\newcommand{\FlattenBlock}[8]{\ConvBlock{#1}{#2}{#3}{#4}{#5}{#6}{#7}{#8}}

\newcommand{\EmissionBox}[6]{%
    \node[draw, thick, fill=blue!5, minimum width=1.2cm, minimum height=1.5cm]
        at (#2-0.2,#3+4.5,#4) {};
    \node[draw, thick, fill=blue!10, align=center, minimum width=1.2cm, minimum height=1.5cm]
        (#1) at (#2,#3,#4) {%
            #5\\%
            {\scriptsize $\lambda #6$}\\%
            $2n_{\rm b}$%
        };
}

\newcommand{\arrowY}{10}
\newcommand{\arrowZ}{5}
\newcommand{\xsp}{1.4}
\newcommand{\xsmall}{1.1}
\newcommand{\xcoord}{0}
\newcommand{\TextOffset}{16}
\newcommand{\BlockOffset}{0.15}

\newcommand{\DimLabel}[7]{%
    \node[
        anchor=south,
        font=\small\bfseries,
        align=center
    ] at
    (#1 + 0.5*#4,
     #2 + #5 + 8,
     #3 + 0.5*#6)
    {#7};
}

\title{\boldmath Lyman Break Galaxy selection and redshift measurement with supervised contrastive learning}

\author[1,2,3]{J.~Choppin de Janvry\,\orcidlink{0009-0008-8066-446X}}
\author[1]{C.~Y\`{e}che\,\orcidlink{0000-0001-5146-8533}}
\author[4]{Arjun~Dey\,\orcidlink{0000-0002-4928-4003}}
\author[1]{C.~Magneville}
\author[1]{C.~Payerne\,\orcidlink{0000-0002-1818-929X}}
\author[5]{J.~Aguilar}
\author[6]{S.~Ahlen\,\orcidlink{0000-0001-6098-7247}}
\author[1]{E.~Armengaud\,\orcidlink{0000-0001-7600-5148}}
\author[5]{S.~Bailey\,\orcidlink{0000-0003-4162-6619}}
\author[7]{F.~Beutler\,\orcidlink{0000-0003-0467-5438}}
\author[8,9]{D.~Bianchi\,\orcidlink{0000-0001-9712-0006}}
\author[10]{D.~Brooks}
\author[11,12]{A.~Carnero Rosell\,\orcidlink{0000-0003-3044-5150}}
\author[5]{E.~Chaussidon\,\orcidlink{0000-0001-8996-4874}}
\author[5]{T.~Claybaugh}
\author[5]{A.~Cuceu\,\orcidlink{0000-0002-2169-0595}}
\author[13]{K.~S.~Dawson\,\orcidlink{0000-0002-0553-3805}}
\author[14]{A.~de la Macorra\,\orcidlink{0000-0002-1769-1640}}
\author[1]{A.~de~Mattia\,\orcidlink{0000-0003-0920-2947}}
\author[10]{P.~Doel}
\author[15,16]{A.~Font-Ribera\,\orcidlink{0000-0002-3033-7312}}
\author[17,18]{J.~E.~Forero-Romero\,\orcidlink{0000-0002-2890-3725}}
\author[19,20,21]{E.~Gazta\~{n}aga\,\orcidlink{0000-0001-9632-0815}}
\author[22]{Satya~{Gontcho A Gontcho}\,\orcidlink{0000-0003-3142-233X}}
\author[23]{G.~Gutierrez}
\author[24,1]{H.~K.~Herrera-Alcantar\,\orcidlink{0000-0002-9136-9609}}
\author[25,26,27]{K.~Honscheid\,\orcidlink{0000-0002-6550-2023}}
\author[28]{M.~Ishak\,\orcidlink{0000-0002-6024-466X}}
\author[4]{S.~Juneau\,\orcidlink{0000-0002-0000-2394}}
\author[29,30]{T.~Karim\,\orcidlink{0000-0002-5652-8870}}
\author[31]{D.~Kirkby\,\orcidlink{0000-0002-8828-5463}}
\author[5]{A.~Kremin\,\orcidlink{0000-0001-6356-7424}}
\author[5]{A.~Lambert}
\author[5]{M.~Landriau\,\orcidlink{0000-0003-1838-8528}}
\author[32]{L.~Le~Guillou\,\orcidlink{0000-0001-7178-8868}}
\author[33,16]{M.~Manera\,\orcidlink{0000-0003-4962-8934}}
\author[4]{A.~Meisner\,\orcidlink{0000-0002-1125-7384}}
\author[15,16]{R.~Miquel}
\author[20]{S.~Nadathur\,\orcidlink{0000-0001-9070-3102}}
\author[34,35]{G.~Niz\,\orcidlink{0000-0002-1544-8946}}
\author[36,37]{E.~Paillas\,\orcidlink{0000-0002-4637-2868}}
\author[1,5]{N.~Palanque-Delabrouille\,\orcidlink{0000-0003-3188-784X}}
\author[38,39,40]{W.~J.~Percival\,\orcidlink{0000-0002-0644-5727}}
\author[41]{F.~Prada\,\orcidlink{0000-0001-7145-8674}}
\author[42]{I.~P\'erez-R\`afols\,\orcidlink{0000-0001-6979-0125}}
\author[43]{C.~Ravoux\,\orcidlink{0000-0002-3500-6635}}
\author[44]{G.~Rossi}
\author[45]{R.~Ruggeri\,\orcidlink{0000-0002-0394-0896}}
\author[46]{E.~Sanchez\,\orcidlink{0000-0002-9646-8198}}
\author[47]{C.~Saulder\,\orcidlink{0000-0002-0408-5633}}
\author[5]{D.~Schlegel}
\author[48,49]{M.~Schubnell}
\author[50]{H.~Seo\,\orcidlink{0000-0002-6588-3508}}
\author[5]{J.~Silber\,\orcidlink{0000-0002-3461-0320}}
\author[21,12]{M.~Siudek\,\orcidlink{0000-0002-2949-2155}}
\author[49]{G.~Tarl\'{e}\,\orcidlink{0000-0003-1704-0781}}
\author[4]{B.~A.~Weaver}
\author[51]{H.~Zou\,\orcidlink{0000-0002-6684-3997}}

\affiliation[1]{IRFU, CEA, Universit\'{e} Paris-Saclay, F-91191 Gif-sur-Yvette, France}
\affiliation[2]{CentraleSup\'{e}lec, Universit\'{e} Paris-Saclay, 3 Rue Joliot Curie, 91190 Gif-sur-Yvette, France}
\affiliation[3]{Department of Physics, University of California, Berkeley, CA 94720, USA}
\affiliation[4]{NSF NOIRLab, 950 N. Cherry Ave., Tucson, AZ 85719, USA}
\affiliation[5]{Lawrence Berkeley National Laboratory, 1 Cyclotron Road, Berkeley, CA 94720, USA}
\affiliation[6]{Department of Physics, Boston University, 590 Commonwealth Avenue, Boston, MA 02215 USA}
\affiliation[7]{Institute for Astronomy, University of Edinburgh, Royal Observatory, Blackford Hill, Edinburgh EH9 3HJ, UK}
\affiliation[8]{Dipartimento di Fisica ``Aldo Pontremoli'', Universit\`a degli Studi di Milano, Via Celoria 16, I-20133 Milano, Italy}
\affiliation[9]{INAF-Osservatorio Astronomico di Brera, Via Brera 28, 20122 Milano, Italy}
\affiliation[10]{Department of Physics \& Astronomy, University College London, Gower Street, London, WC1E 6BT, UK}
\affiliation[11]{Departamento de Astrof\'{\i}sica, Universidad de La Laguna (ULL), E-38206, La Laguna, Tenerife, Spain}
\affiliation[12]{Instituto de Astrof\'{\i}sica de Canarias, C/ V\'{\i}a L\'{a}ctea, s/n, E-38205 La Laguna, Tenerife, Spain}
\affiliation[13]{Department of Physics and Astronomy, The University of Utah, 115 South 1400 East, Salt Lake City, UT 84112, USA}
\affiliation[14]{Instituto de F\'{\i}sica, Universidad Nacional Aut\'{o}noma de M\'{e}xico, Circuito de la Investigaci\'{o}n Cient\'{\i}fica, Ciudad Universitaria, Cd. de M\'{e}xico C.~P.~04510, M\'{e}xico}
\affiliation[15]{Instituci\'{o} Catalana de Recerca i Estudis Avan\c{c}ats, Passeig de Llu\'{\i}s Companys, 23, 08010 Barcelona, Spain}
\affiliation[16]{Institut de F\'{i}sica d'Altes Energies (IFAE), The Barcelona Institute of Science and Technology, Edifici Cn, Campus UAB, 08193, Bellaterra (Barcelona), Spain}
\affiliation[17]{Departamento de F\'isica, Universidad de los Andes, Cra. 1 No. 18A-10, Edificio Ip, CP 111711, Bogot\'a, Colombia}
\affiliation[18]{Observatorio Astron\'omico, Universidad de los Andes, Cra. 1 No. 18A-10, Edificio H, CP 111711 Bogot\'a, Colombia}
\affiliation[19]{Institut d'Estudis Espacials de Catalunya (IEEC), c/ Esteve Terradas 1, Edifici RDIT, Campus PMT-UPC, 08860 Castelldefels, Spain}
\affiliation[20]{Institute of Cosmology and Gravitation, University of Portsmouth, Dennis Sciama Building, Portsmouth, PO1 3FX, UK}
\affiliation[21]{Institute of Space Sciences, ICE-CSIC, Campus UAB, Carrer de Can Magrans s/n, 08913 Bellaterra, Barcelona, Spain}
\affiliation[22]{University of Virginia, Department of Astronomy, Charlottesville, VA 22904, USA}
\affiliation[23]{Fermi National Accelerator Laboratory, PO Box 500, Batavia, IL 60510, USA}
\affiliation[24]{Institut d'Astrophysique de Paris. 98 bis boulevard Arago. 75014 Paris, France}
\affiliation[25]{Center for Cosmology and AstroParticle Physics, The Ohio State University, 191 West Woodruff Avenue, Columbus, OH 43210, USA}
\affiliation[26]{Department of Physics, The Ohio State University, 191 West Woodruff Avenue, Columbus, OH 43210, USA}
\affiliation[27]{The Ohio State University, Columbus, 43210 OH, USA}
\affiliation[28]{Department of Physics, The University of Texas at Dallas, 800 W. Campbell Rd., Richardson, TX 75080, USA}
\affiliation[29]{Center for Astrophysics $|$ Harvard \& Smithsonian, 60 Garden Street, Cambridge, MA 02138, USA}
\affiliation[30]{Department of Astronomy \& Astrophysics, University of Toronto, Toronto, ON M5S 3H4, Canada}
\affiliation[31]{Department of Physics and Astronomy, University of California, Irvine, 92697, USA}
\affiliation[32]{Sorbonne Universit\'{e}, CNRS/IN2P3, Laboratoire de Physique Nucl\'{e}aire et de Hautes Energies (LPNHE), FR-75005 Paris, France}
\affiliation[33]{Departament de F\'{i}sica, Serra H\'{u}nter, Universitat Aut\`{o}noma de Barcelona, 08193 Bellaterra (Barcelona), Spain}
\affiliation[34]{Departamento de F\'{\i}sica, DCI-Campus Le\'{o}n, Universidad de Guanajuato, Loma del Bosque 103, Le\'{o}n, Guanajuato C.~P.~37150, M\'{e}xico}
\affiliation[35]{Instituto Avanzado de Cosmolog\'{\i}a A.~C., San Marcos 11 - Atenas 202. Magdalena Contreras. Ciudad de M\'{e}xico C.~P.~10720, M\'{e}xico}
\affiliation[36]{Instituto de Estudios Astrof\'isicos, Facultad de Ingenier\'ia y Ciencias, Universidad Diego Portales, Av. Ej\'ercito Libertador 441, Santiago, Chile}
\affiliation[37]{Steward Observatory, University of Arizona, 933 N. Cherry Avenue, Tucson, AZ 85721, USA}
\affiliation[38]{Department of Physics and Astronomy, University of Waterloo, 200 University Ave W, Waterloo, ON N2L 3G1, Canada}
\affiliation[39]{Perimeter Institute for Theoretical Physics, 31 Caroline St. North, Waterloo, ON N2L 2Y5, Canada}
\affiliation[40]{Waterloo Centre for Astrophysics, University of Waterloo, 200 University Ave W, Waterloo, ON N2L 3G1, Canada}
\affiliation[41]{Instituto de Astrof\'{i}sica de Andaluc\'{i}a (CSIC), Glorieta de la Astronom\'{i}a, s/n, E-18008 Granada, Spain}
\affiliation[42]{Departament de F\'isica, EEBE, Universitat Polit\`ecnica de Catalunya, c/Eduard Maristany 10, 08930 Barcelona, Spain}
\affiliation[43]{Universit\'{e} Clermont-Auvergne, CNRS, LPCA, 63000 Clermont-Ferrand, France}
\affiliation[44]{Department of Physics and Astronomy, Sejong University, 209 Neungdong-ro, Gwangjin-gu, Seoul 05006, Republic of Korea}
\affiliation[45]{Queensland University of Technology, School of Chemistry \& Physics, George St, Brisbane 4001, Australia}
\affiliation[46]{CIEMAT, Avenida Complutense 40, E-28040 Madrid, Spain}
\affiliation[47]{Max Planck Institute for Extraterrestrial Physics, Gie\ss enbachstra\ss e 1, 85748 Garching, Germany}
\affiliation[48]{Department of Physics, University of Michigan, 450 Church Street, Ann Arbor, MI 48109, USA}
\affiliation[49]{University of Michigan, 500 S. State Street, Ann Arbor, MI 48109, USA}
\affiliation[50]{Department of Physics \& Astronomy, Ohio University, 139 University Terrace, Athens, OH 45701, USA}
\affiliation[51]{National Astronomical Observatories, Chinese Academy of Sciences, A20 Datun Road, Chaoyang District, Beijing, 100101, P.~R.~China}

\emailAdd{jean.choppindejanvrydev@gmail.com}
\emailAdd{christophe.yeche@cea.fr}

\abstract{Some of the next steps for high-precision cosmology lie within the high-redshift, high-density universe. Spectroscopic survey experiments such as the Dark Energy Spectroscopic Instrument (DESI)'s second phase DESI Run 2 will shift towards probing Lyman Break Galaxy (LBG) populations from $z\sim2$ to $z\sim4.5$. For this faint sample, spectroscopic redshift measurement and sample decontamination remains a challenge, even after target selection. We propose an approach based on supervised weighted contrastive learning, in order to both learn a redshift representation for spectra and decontaminate the sample from quasars and low redshift emission line galaxies. This strategy  generalizes the contrastive learning loss approach with continuous relationship weights, such that the network simultaneously learns redshift and classification tasks. The model shows stronger outlier classification and comparable redshift identification performances when compared to the previous network used for DESI (a modified version of \qnet) on the same dataset. In particular, contrastive learning is well suited to the small, visually-inspected sample used for training and testing, especially given the multi-task nature of this work.}

\begin{document}
\maketitle
\flushbottom


\section{Introduction}\label{sec:intro}

Populations of galaxies in the high-redshift universe ($z\gtrsim2$) remain relatively unexplored, especially in the context of large-scale, multiplexed spectroscopic redshift surveys. In that regard, on-going, upcoming and proposed Stage IV and Stage V surveys such as the Dark Energy Spectroscopic Instrument (DESI) Run 2 \cite{schlegel2022spectroscopicroadmapcosmic}, Spec-S5 \cite{besuner2025spectroscopicstage5experiment, schlegel2022spectroscopicroadmapcosmic}, MegaMapper \cite{schlegel2022megamapper}, the Subaru Prime Focus Spectrograph (PFS) \cite{Takada_2014_PFS, Tamura_2016_PFS}, the Wide field Spectroscopic Telescope (WST) \cite{bacon2024wstwidefieldspectroscopic}, the Maunakea Spectroscopic Explorer (MSE) \cite{themsescienceteam2019detailedsciencecasemaunakea}, 4-metre Multi-Object Spectroscopic Telescope (4MOST) \cite{deJong_2019_4MOST} among others have expressed scientific interest in targeting the population of Lyman Break Galaxies (LBGs). 

LBGs are star-forming galaxies of the early universe (above $z\gtrsim1.5$). Due to absorption by neutral Hydrogen, their spectra can be characterized by a flux break at wavelengths shorter than the Lyman limit ($912 \angstrom$ at rest frame) and a flux drop shortward of the Lyman-$\alpha$ ($\lya$) spectral feature ($1216 \angstrom$ at rest frame) \cite{Steidel1998_LBGs, GiavaliscoLBGs2002, Shapley2003restframeLBGs, Ly_2009_Subaru_GALEX_LBG, Shapley_2011_properties_z2_4, Forster_Schreiber_2020_SFGalaxies_CosmicNoon}. 
The $\lya$ line itself can be both in emission or absorption. A dense population of galaxies at high-redshift offers promising precision cosmology measurements and astrophysical science. For instance, the LBG sample will be particularly useful for measurements of primordial non-gaussianity \cite{payerne2025forecastingprimordialnongaussianityunions}, since this measurement is currently only performed at that redshift range using the quasar (QSO) $\lya$ forest at $z\gtrsim2.1$ in DESI \cite{QSO.TS.Chaussidon.2023, ChaussidonY1fnl}. Moreover, LBGs will serve as tomographic probes of the line of sight with absorption of Hydrogen in their $\lya$ forest \cite{Herrera_Alcantar_2025_LBGs_Lyaforest}, enabling creation of void catalogs \cite{Contarini2026_voids}. The LBG galactic environment further paves the way for better understandings of the circum-galactic medium (CGM) through the relationships between intergalactic gas and galaxies \cite{Chung_2019_CGM_LBGs}, and LBGs have been identified as tracers for the ionizing spectrum to better understand the epoch of reionization \cite{Steidel2018LBGsionizing}. These encouraging paths for new high-precision measurements and physical insights incite future spectroscopic surveys to observe the LBG population, and once the Dark Energy Spectroscopic Instrument (DESI) concludes its planned eight-year survey of $\sim17.000$ sq.deg$^2$ of the sky, the continuation DESI Run 2 \cite{schlegel2022spectroscopicroadmapcosmic} will commence and plans to dedicate observation programs to the LBGs. 

Before DESI Run 2 starts, DESI will have measured $\sim63$ million spectroscopically confirmed redshifts of galaxies and quasars. The instrument itself \cite{DESI2022.KP1.Instr, DESI2016b.Instr} operates on the Mayall $4$-meter telescope at Kitt Peak National Observatory and can observe up to 5000 objects at once \cite{SurveyOps.Schlafly.2023} over a $\sim3^\circ$ field of the sky with fiber-fed robotic positioners \cite{Corrector.Miller.2023, FiberSystem.Poppett.2024}. The instrument used for DESI Run 2 plans to remain broadly similar, though some noteworthy upgrades to the spectroscopic system will be included, especially in the blue channel of the spectrographs. This is particularly relevant given the $\lya$ $1216$ feature and Lyman break [912 \AA] appear in the blue spectrograph range ($3600-5930$ \AA \cite{Spectro.Pipeline.Guy.2023}) for most of the LBG redshift range observed ($z\gtrsim2$). The spectra observed are reduced through DESI's spectroscopic pipeline \cite{Spectro.Pipeline.Guy.2023} and the first data release (DR1), containing more than 18 million unique spectroscopic redshifts, is publicly available \cite{DESI2024.I.DR1}. Moreover, DESI DR1 and DR2 results on the nature of dark energy \cite{DESI.DR2.BAO.cosmo, DESI2024.VII.KP7B} strongly motivate further studies of the high-redshift universe. DESI Run 2 will reinforce the populations already observed by DESI, support other experiments such as the Vera C. Rubin telescope \cite{LSST.Ivezic.2019} through dedicated programs, and probe new galaxy populations such as the LBGs, which are expected to exceed the density of DESI QSO targets \cite{QSO.TS.Chaussidon.2023} by at least an order of magnitude \cite{RuhlmannKleider2024_LBGs}. However, due to scarce easily identifiable spectroscopic features besides potential $\lya$ emission or absorption, LBGs require longer integration times ($\sim120$min) than QSOs to confidently measure their redshift and decontaminate the sample at the target densities necessary for DESI Run 2.

At redshifts $z\gtrsim2$, $\lya$ feature enters the optical wavelengths available for the DESI spectrographic range ($\sim3600-9800 \angstrom$ \cite{DESI2016b.Instr}), thereby imposing a lower bound on the observable redshift range. The upper redshift limit of the population is around $z\sim4.5$, imposed by the target selection \cite{SelectionLBG_highz_Payerne2025}. Target selection of LBGs is done through color dropout techniques\footnote{Or associated neural networks. For selection of high redshift galaxies, Random Forests are often used \cite{QSO.TS.Chaussidon.2023, SelectionLBG_highz_Payerne2025}.} in which broadband imaging is used (\textit{ugri} in the case of DESI). Color dropout detects a high magnitude difference due to the presence of the Lyman break. In the case of LBGs, u-g color is used, so at $z\gtrsim4.5$, dropout selection does not obtain LBGs since the break shifts out of the broadband selection. However, target selection has its own share of limitations, since contaminants also leak into the sample. In the case of LBGs, the sample is contaminated by low redshift ($z\lesssim0.15$) emission line galaxies (ELG) where the [\ion{O}{2}] \wl{3726,3729} doublet is confused with \lya \wl{1216} emission. Another important contaminant in LBG selection is the population of high-redshift QSOs ($z\gtrsim2$), whose broadband photometric properties can mimic those of LBGs. However, in contrast to LBGs, QSOs typically exhibit broad emission lines with large full width at half maximum (FWHM) values, originating from high-velocity gas in the broad-line region surrounding the central supermassive black hole. Their spectra also commonly show active galactic nuclei (AGN) high-ionization emission lines, such as high-ionization lines (\ion{C}{4} \wl{1549}, \ion{Si}{4} \wl{1393,1402}), produced by gas photo-ionized by radiation from the accretion disk. These contaminants inevitably survive the selection cuts of the LBG sample, at the $\sim10\%$ level each in the observed sample (dependent on choices of target density and selection methods). 
DESI aims to observe LBG targets at a surface density of $\sim900\,\mathrm{deg^{-2}}$, achieving around $\sim2$ million confident redshifts as a baseline for the DESI Run 2 science case. 
In the case of LBGs, DESI Run 2 expects to recover $\sim400\,\mathrm{deg^{-2}}$ secure spectroscopic redshifts out of the $\sim900$ observed. Therefore, it is crucial to improve the redshift estimation after observations in order to increase the completeness of the observed LBG sample. To do so, machine learning techniques have been previously employed in DESI. For LBGs, the current approach is derived from \qnet \cite{busca2018_quasarnet}. \qnet is also used in DESI QSO redshift measurements \cite{QSO.TS.Chaussidon.2023}, and was adapted specifically to LBG spectra and presented in \cite{RuhlmannKleider2024_LBGs}. In order to distinguish from \qnet, the variant of the network suited to LBGs is named \lbgnet throughout this study. The architecture developed in this work is named \zlbg and aims at providing an alternative machine learning pipeline to increase the sample quality and redshift performance of the LBG sample. \zlbg leverages contrastive learning to predict redshift and classify LBG spectra in order to decontaminate the observed sample. 

The following Section~\ref{sec:motivation} motivates the use of contrastive learning in this context, and further presents the LBG galaxies of DESI. Section~\ref{sec:architecture} presents the supervised contrastive learning methodology employed and introduces the network architecture of both pipelines. Then, Section~\ref{sec:data} presents the construction of the test and training datasets with the help of visual inspection data campaigns and data augmentation. Network performance, results and stress tests are displayed in Section~\ref{sec:results}. Finally, we conclude this study in Section~\ref{sec:conclusion}. Appendix~\ref{sec:appendix:redshift_weighting} and Appendix~\ref{sec:appendix:class_weighting} describe further stress tests and parameter choices for the \zlbg pipeline, while Appendix~\ref{sec:appendix:lz} explores prior choices for spectroscopic measurements after selection by the networks. Finally, Appendix~\ref{sec:appendix:bad_spectra} explores treatment of low signal-to-noise ratio (SNR) spectra by the pipeline, where no confident redshift can be assessed.
\section{Motivation and context}\label{sec:motivation}

A key motivation to enhance the catalog quality of the LBG sample for DESI is to reinforce the decontamination and redshift inference at the spectral level, after observations. Spectroscopic redshift measurement is in general straight-forward for most $z<1.6$ galaxies: the \redrock pipeline \cite{Redrock.Bailey.2024} uses templates of multiple emission and absorption lines, exploring the designated redshift range and iterating over redshifts to minimize the $\chi^2$.  
However, for faint, high-redshift targets such as LBGs \cite{QSO.TS.Chaussidon.2023}, catastrophic outliers (such as low redshift ELGs \cite{ELG.TS.Raichoor.2023} or QSOs \cite{QSO.TS.Chaussidon.2023}) in redshift contaminate the sample if attempting to fit over a naive, broad redshift range. LBG sources are fainter, thus have low signal-to-noise ratios (SNR), and there are many confusing emission line detections due to the large redshift range to explore. Therefore, a prior on the redshift is useful to constrain the explored redshift range before using \redrock. For LBGs \cite{RuhlmannKleider2024_LBGs}, a uniform prior $\mathcal{U}(z_\mathrm{pred}-\lz, z_\mathrm{pred}+\lz)$ is used on \redrock centered on the predicted redshift $z_\mathrm{pred}$, with $\lz=0.1$ fixed. The predicted redshift $z_{\mathrm{pred}}$ is obtained leveraging machine learning: for QSOs, the DESI pipeline currently uses the \qnet algorithm \cite{busca2018_quasarnet}. More details about the specific flowchart of DESI quasar processing can be found in Figure 9 of \cite{QSO.TS.Chaussidon.2023}, and further enhancements to \qnet measurements are presented in \cite{Green.ActiveLearning.QSO.2025}. The current state of the art for LBG processing in DESI is described in \cite{RuhlmannKleider2024_LBGs} and uses a modified version of \qnet (named \lbgnet in the context of this work and further presented in Section~\ref{sec:architecture}) that includes LBG-specific spectra templates and is trained to recognize LBG spectra, decontaminate the sample from outliers and suggest a point estimate redshift $z_\mathrm{pred}$. This work shares the same goal and will therefore evaluate the same metrics as \cite{RuhlmannKleider2024_LBGs}. However, motivated by recent improvements to \qnet priors for \redrock in the context of QSO redshift measurements, the criteria for redshift prediction success will now depend on redshift, and \redrock performance will be tested with $\lz=0.025(1+z_\mathrm{pred})$ prior. The amplitude pre-factor $0.025$ is chosen such that $z_\mathrm{pred}=3$ matches the fixed $\lz=0.1$ criteria from \lbgnet \cite{RuhlmannKleider2024_LBGs}. 

The software pipeline\footnote{Software is available at: \href{https://github.com/JeanCHDJdev/desi-lbg}{\textcolor{blue}{github.com/JeanCHDJdev/desi-lbg}}} developed in this work is named \zlbg and uses supervised contrastive learning. 
Contrastive learning (CL) is a recently developed machine learning approach, both in the unsupervised \cite{chen2020simpleframeworkcontrastivelearning} and supervised \cite{khosla2021_supervisedcontrastivelearning} paradigms. CL, has since known many applications in astrophysics and cosmology, for example in the case of galaxy physics \cite{Martinez_Solaeche_2024_eCALIFA_CL}, strong lensing \cite{Cheng2020_stronglenses_unsupervised, Hayat2021_selfsupervisedrepresentation, Stein2022_stronggravlenses_selfsupervised}, gravitational wave detection and classification \cite{2026.Wang.GW.SupervisedCL} or omnimodal foundation models \cite{Parker_2024_AstroCLIP, Parker.2025.AION}. A more thorough overview of some of the application cases of CL in astrophysics is described in \cite{huertascompany2023briefreviewcontrastivelearning}. 
CL maps data into an embedding space in which samples with the same labels are placed close together, while samples with different labels are pushed farther apart. This encourages the learned representation to capture the features that are most informative for distinguishing between the target labels. In this work, similarity is defined by both object classification and redshift. Consequently, the embedding is encouraged to emphasize features relevant to these targets while dismissing object properties that do not contribute directly to predicting the object class or redshift, such as color or metal absorption lines. Once the vector space has been learnt, the representations can be used for a variety of downstream tasks.

\begin{figure}[h]
    \centering
    \includegraphics[width=0.92\textwidth]{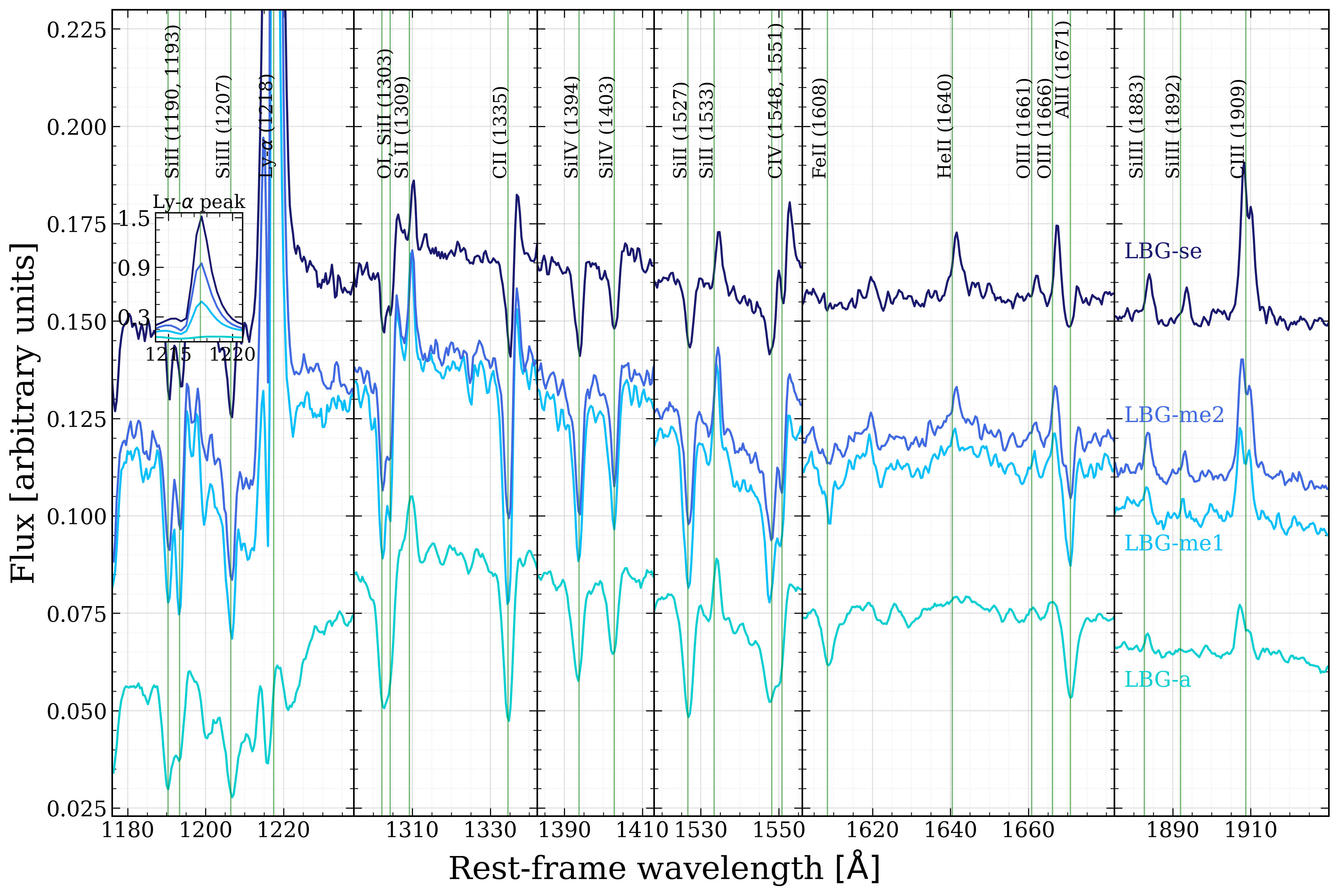}
    \caption{Highlighting some physical properties differences between the different types of LBGs at different wavelength ranges. The most noticeable feature is the \lya line (inset plot due to being out of frame). However, as discussed in text, amplitudes of other lines also vary alongside physical properties. Spectras are offset on the flux axis for readability. From top to bottom: LBGse, LBGme-2, LBGme-1 and LBGa.}
    \label{fig:motivation:zoom_features}
\end{figure}

A CL architecture is well motivated in this case. First, the network builds over few galaxy classes. LBGs are discriminated by their characteristic $\lya$ intensity: LBGa (LBG-absorption), LBGme (LBG-medium emission) and LBGse (LBG-strong emission), roughly following earlier existing classifications of LBGs \cite{Shapley2003restframeLBGs}. In particular, the LBGse are characterized by a very strong $\lya$ emission and are often known as $\lya$ emitters (LAE). While LAEs present the Lyman break feature and can therefore fall into LBG selections (here named LBGse), specific target selection geared towards LAE targets in DESI Run 2 will probe an additional $\sim2.2$ million LAEs with a $30$ min exposure time for high-redshift science. DESI's LAE program will perform target selection using the Intermediate Band Imaging Survey (IBIS) \cite{Ebina_2026_Clustering_MB_IBIS, ebina20263dclusteringlymanalpha} and medium-band flux excess to detect $\lya$ emission, instead of looking for the Lyman break. In general, LBGse selected by a dropout technique (LBG-type selection, using broadband filters) are brighter in broadband magnitude than LAEs selected by medium band imaging, due to the presence of a brighter continuum. The LBGse denomination is conserved due to the selection method, which focused on the Lyman break rather than $\lya$ emission. Figure~\ref{fig:motivation:zoom_features} displays stacked spectras per different LBG sub-types in the UV rest-frame, as well as common emission and absorption lines. The LBGme class contains two templates at varying $\lya$ intensity, named LBGme-1 and LBGme-2. We present in detail the construction of these templates in Section~\ref{sec:data:templates}. The aforementioned figure should be compared to Table 3 presented by \cite{Shapley2003restframeLBGs}, where the evolution of rest-frame equivalent widths for some common low ionization interstellar lines (LIS: \ion{Si}{2} \wl{1260}, \ion{O}{1}+\ion{Si}{2} \wl{1303}, \ion{Si}{2} \wl{1526}, \ion{C}{2} \wl{1335}, \ion{Fe}{2} \wl{1608}, \ion{Al}{2} \wl{1670}), high-ionization lines (\ion{Si}{4} \wl{1393,1402}, \ion{C}{4} \wl{1549}), and nebular emission lines (NE: [\ion{O}{3}] \wl{1661,1666}, \ion{C}{3}] \wl{1909}) are shown for different LBG types (separated by their $\lya$ equivalent width). In LBGa spectra, there is stronger absorption compared to LBGme and LBGse, whereas LBGse presents stronger nebular emission lines. LIS are well absorbed in LBGa, since LBGa presents continuous, dense interstellar medium environments with larger E(B-V). One other example to showcase is the progressive disappearance of nebular line \ion{He}{2} \wl{1640} as the sample goes from LBGse to LBGa. 
The templates presented in Figure~\ref{fig:motivation:zoom_features} are similar to Figure 12 of \cite{RuhlmannKleider2024_LBGs}. Nonetheless, some differences are to be noted. These differences are discussed in Section~\ref{sec:data:templates}.

Choosing CL is also motivated by the small size of the training set. While \qnet originally leveraged a large dataset of QSOs for training (249,762 unique QSOs \cite{busca2018_quasarnet} from the Baryon Oscillation Spectroscopic Survey (BOSS) \cite{BOSS.GilMartin.2020}), \lbgnet cannot leverage the same amount of data due to lackluster confident LBG observations since samples are yet to be observed. 

However, over the course of the nominal DESI survey, multiple dedicated programs have taken place in preparation for DESI Run 2, often known as \textit{pilot programs}\footnote{Also mentioned as \textit{tertiary programs} in other DESI works.}. These include testing target selection functions for LBG/LAE tracers in different fields, hence giving various sample populations to test and train from. 
Out of the observed samples, characterizing LBG populations (by type and redshift) and assessing their quality is done through time and resource intensive Visual Inspection (VI) campaigns, where 3 reviewing observers independently measure redshift, type and confidence of LBG spectra. For DESI, the VI procedure initially tackled during survey validation (SV) of the instrument is described in \cite{VIQSO.Alexander.2023}. This VI procedure acts as our ground truth catalog for both testing and training. The aforementioned catalog of QSOs built during SV will serve as a contaminant dataset for our LBGs. Specifically, we use three VI datasets built over the DESI Run 1 observation years. The first dataset, named \setzero, is a complete visual inspection of a target selection. Most of this dataset is used for training, as detailed in \ref{sec:data:training_testing}, and a chunk is reserved for validation and testing. The second dataset, \setone, is exclusively used for testing, and is also provided from LBG pilot programs. We only take the identified LBGs from this dataset, and it comes from the combined VI of some spectra from other pilot programs. Finally, we use \settwo: this set contains specific ELG and QSO contaminants that were visually inspected during survey validation (SV) of DESI, as detailed in \cite{VIQSO.Alexander.2023, VIGalaxies.Lan.2023}. These ELGs and QSOs serve as representative populations of possible contaminants during target selection. Additionally, we also conserve ELGs and QSOs observed and VIed in \setzero.
A full description of the tiles, co-adds, pilot program origins and productions used are described on \href{https://doi.org/10.5281/zenodo.21775480}{\textcolor{blue}{\texttt{zenodo}}}, alongside the data to recreate the figures for this work. Alternatively, \setzero is described in Table 5 of \cite{RuhlmannKleider2024_LBGs}.

This motivates the use of CL, which can learn meaningful representations of the spectra out of very small datasets, especially if classes are balanced inside datasets with contaminant over-representation. In particular, CL highly benefits from data augmentations over noisy and sparse datasets, as the network is built to be insensitive to certain variations (for example, noise, shape, galaxy type...). Finally, the multi-task nature of the work (contaminant selection / redshift identification) justifies obtaining meaningful representations rather than single tasks with a decoder: see e.g. \cite{Hayat2021_selfsupervisedrepresentation} for examples of multi-task astronomical image classification and regression after CL representation training. Thus, CL is a well suited architecture for this problem. In the following Section~\ref{sec:architecture}, we present the architecture of \zlbg and \lbgnet as well as the CL infrastructure.
\section{Architecture}\label{sec:architecture}


In this section, we present the architectural choices of the \zlbg pipeline (this work) and compare them to the modified \qnet architecture described in \cite{RuhlmannKleider2024_LBGs}. As specified in the introduction, we hereby refer to this modified architecture as \lbgnet, in order to differentiate from the original version of \qnet \cite{busca2018_quasarnet}. We first describe the common backbone encoder in Section \ref{sec:architecture:encoder}, then discuss the contrastive loss function and the associated weighting necessary in Section \ref{sec:architecture:loss} and Section \ref{sec:architecture:weights}. Finally, we transform the obtained representations into classified spectra and inferred redshifts with downstream tasks in Section \ref{sec:architecture:classification} and \ref{sec:architecture:z_inference}. 

\subsection{Encoder}\label{sec:architecture:encoder}

\begin{figure}[h]
\centering
\input{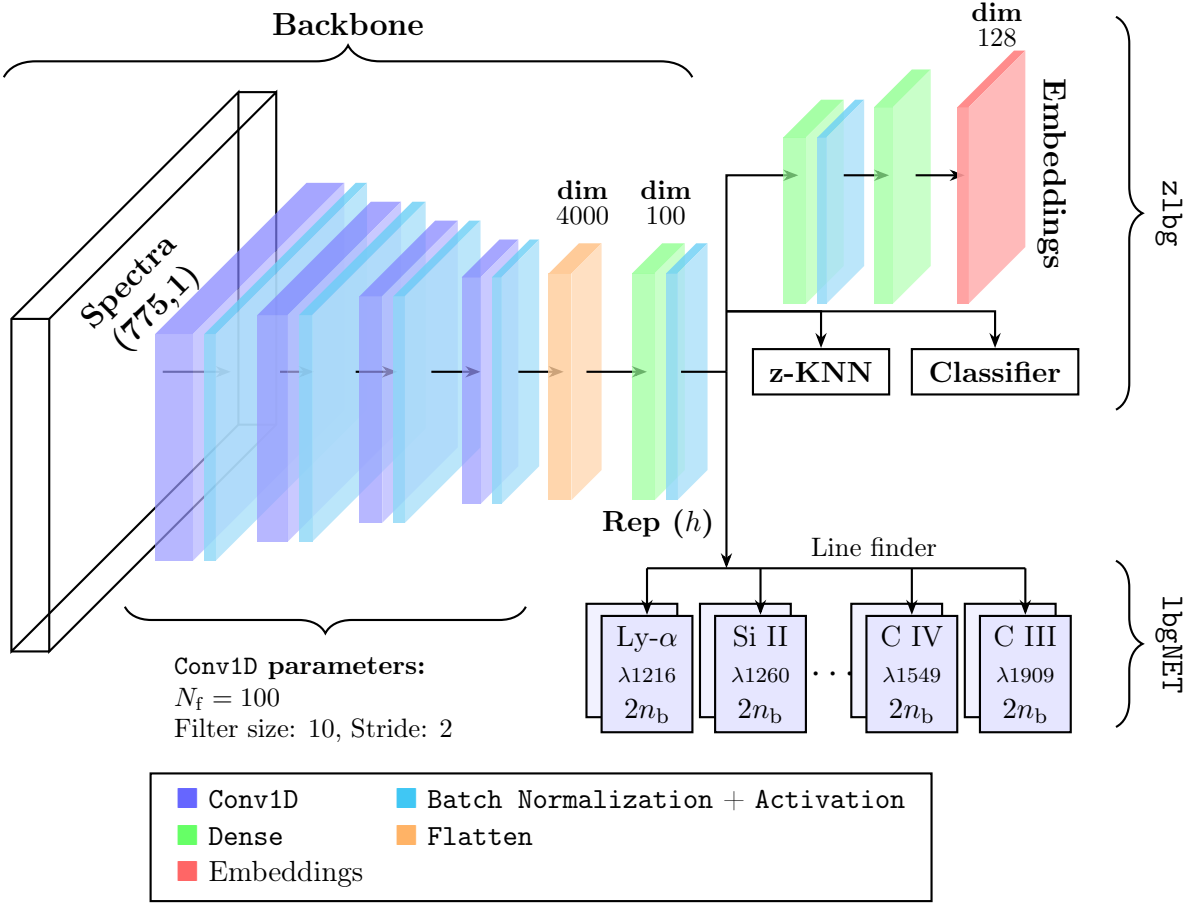}
\caption{Architecture schematic of the \zlbg (top right) and \lbgnet (bottom right) networks. The details of implementation are described in text. The encoder shares the same backbone as \qnet (up to the representation layer $\mathbf{Rep}$, which is the encoder), such that the only difference in performance can be attributed to the loss function. The \texttt{Flatter}, representation ($\mathbf{Rep}$) and embedding ($\mathbf{Proj}$) layers have their dimensions specified. The convolution layers ingesting the spectra contain $N_f=100$ filters each with a kernel size of $10$ and $2$ strides.}
\label{fig:architecture}
\end{figure}

The backbone network architecture for both \lbgnet and \zlbg is summarized in Figure~\ref{fig:architecture}, and is conserved from the original \qnet architecture. The input spectra is linearly binned to dimension (775,1) ($\mathrm{d}\lambda=8\angstrom$) using per pixel inverse variance. We further discuss pre-processing of the spectra in Section~\ref{sec:data}. The encoder itself consists of 4 layers of convolution \texttt{Conv1D}. Each layer is followed by a \texttt{BatchNormalization} (provoking z-score normalization of the representation for convergence) and a non-linear Rectified Linear Unit (ReLU) \texttt{Activation}. The convolved spectra are then flattened and sent through a \texttt{Dense} layer. After this layer, architectural choices differ between \zlbg, \qnet and \lbgnet. This layer is called the representation layer $\mathbf{Rep}$, with elements $h\in\mathbf{Rep}$, such that spectra $x$ verify $h=\mathbf{Rep}(x)$. 

In \qnet and \lbgnet, the representation elements $h$ are used as final layer embeddings in the "line finder" component, at the end of the network. In the specific case of \lbgnet, for each absorption or emission line one can expect, two groups of $n_{\mathrm{b}}=40$ neurons slice the spectrum into boxes. One group looks for the presence of the line (the "line detector"), while the "position finder" group gives an estimate for the redshift according to that line. In Figure~\ref{fig:architecture}, the \lbgnet architecture is shown after the $\mathbf{Rep}$ layer: two boxes of $n_b$ neurons are assigned to each line, here showing 4 out of the 16 lines included in the network. The spectroscopic lines considered by \lbgnet can be found in Table 3 of \cite{RuhlmannKleider2024_LBGs}. Per spectrum, a confidence level for each of the 16 lines is defined using the "line detector". The only significant architecture difference between \lbgnet and \qnet is the list of spectroscopic lines used as well as the number of neurons in the boxes $n_{\mathrm{b}}$. A spectrum is qualified as an LBG by \lbgnet if the fifth highest confidence level among all lines is above a given confidence threshold $\tau_\zlbg$, as defined by \cite{RuhlmannKleider2024_LBGs}. The redshift found by the "position finder" for the line with the highest confidence becomes the predicted LBG redshift. 

This is not the end of the network for \zlbg. The representation layer elements $h$ are processed by a multi-layer perceptron (MLP) head consisting of two Dense layers projecting onto the contrastive embeddings $z$ of dimension $128$, ending with $L_2$ normalization at the embeddings level. Consequently, $\mathbf{Rep}$ is the representation (encoder) architecture and $\mathbf{Proj}$ is the projection to the latent space, such that for an element $x$ of the spectra, the corresponding embedding $v$ can be written $v=\mathbf{Proj}(\mathbf{Rep}(x))=\mathbf{Proj}(h)$ (red layer in the architecture). The key difference is the loss function, where contrastive learning is involved. We present the contrastive learning loss in the following Section \ref{sec:architecture:loss}. This loss function applies to the $v$ embeddings, but the information for downstream tasks is extracted from the $\mathbf{Rep}$ layer: we follow here recommendations from \cite{chen2020simpleframeworkcontrastivelearning} (Section 4.2) and throw away the projection head after training. This is induced by the loss of information due to the contrastive loss, as it is trained to be invariant to data transformations (e.g. here, galaxies of the same type and/or at the same redshift). By conjecture, the representation layer saves some of this information that the embeddings would in principle remove to align as much as possible with the contrastive loss. Therefore, the underlying $\mathbf{Rep}$ layer holds the information. Dimensionality reduction techniques such as Uniform Manifold Approximation and Projection (UMAP \cite{mcinnes2020umapuniformmanifoldapproximation}), t-SNE \cite{VanDerMaaten.tSNE.2008} or Principal Component Analysis (PCA) can however prove very useful to visualize the embeddings and check they align well with the data augmentations and loss objectives. We show the embeddings represented by UMAP in Section~\ref{sec:results}. As such, classification (Classifier) and redshift measurement (z-KNN) tasks are directly fit on the $\mathbf{Rep}$ layer representations of the training set, as shown in Figure~\ref{fig:architecture}. The downstream tasks for classification and redshift are further described in Sections~\ref{sec:architecture:classification} and \ref{sec:architecture:z_inference}.

\subsection{Loss}\label{sec:architecture:loss}

While the architecture used in \qnet\footnote{\href{https://github.com/ngbusca/QuasarNET}{\textcolor{blue}{github.com/ngbusca/QuasarNET}}} \cite{busca2018_quasarnet} is well proven for quasars and leverages finding multiple lines using the line finder, the limitations of our training dataset described in Section~\ref{sec:motivation} motivate the use of contrastive learning.

The visual inspection (VI) campaigns have nonetheless fully characterized a subset of LBG galaxies in both type and redshift (\setzero). This information is leveraged as labels and therefore motivates a supervised approach to CL. The CL philosophy aims at building a representation in an embedding vector space of the elements (here, galaxy spectra) in a sample by comparing the relationships between the elements. Thus, contrastive methods define "positive" elements with respect to an element $x$ of index $i$ to be the set of elements $P(i)$ that are "similar". This similarity is measured using the cosine distance, measured over the projected embeddings when computing the loss function. Data augmentation provokes similarity over augmented features: as such, for spectra, one possible augmentation $x_\mathrm{aug}$ (of index $i_\mathrm{aug}$) would be changing the Signal-to-Noise Ratio (SNR) of the signal, by resampling the noise. This allows the representation in embedding space to gain invariance with respect to the spectra's SNR, as the pair of spectra is now positive ($i_\mathrm{aug}\in P(i)$). For the purpose of this study the embedding space should only depend on redshift and galaxy type, while the embeddings must be invariant to e.g. SNR, in order to not learn noise features or detector systematics. This will propagate upstream to the representation layer. In practice, a contrastive supervised loss function can be described by the following equation \ref{eq:supcon_loss}:

\begin{equation}\label{eq:supcon_loss}
    \mathcal{L}_{\mathrm{out}}^{\mathrm{sup}}
= \sum_{i \in I} \mathcal{L}_{\text{out}, i}^{\text{sup}}
= \sum_{i \in I}
\frac{-1}{\lvert P(i) \rvert}
\sum_{p \in P(i)}\log\left(
\frac{\exp \left( v_i \cdot v_p / \Theta \right)}
{\sum_{a \in A(i)} \exp \left( v_i \cdot v_a / \Theta \right)}
\right)
\end{equation}
Here, we employ the close notations to the seminal work for supervised contrastive learning \cite{khosla2021_supervisedcontrastivelearning}. This expression is a generalization of the self-supervised contrastive loss function\footnote{The self-supervised contrastive loss works by pairs of augmentations, so there is no need for summing over the positives in the loss.} for a batch of spectra where external label information allows for multiple positives within one batch. We employ $\mathcal{L}_{\mathrm{out}}^{\mathrm{sup}}$ as it reportedly performs better than $\mathcal{L}_{\mathrm{in}}^{\mathrm{sup}}$, another proposed method to generalize the CL loss in \cite{khosla2021_supervisedcontrastivelearning} (equations 2 and 3). In expression \ref{eq:supcon_loss}, $P(i)$ 
is the set of indices of positive elements with respect to the element of index $i$, excluding the latter of the set. Its cardinality is given by $\left|P(i)\right|$. $A(i)=I\setminus\left\{i\right\}$ is the set of indices of the batch $I$ excluding $i$ (designated as anchors in literature). To summarize, $v_p$ is the embedding of a positive element $x_p$ with respect to the element $x$ of index $i$ and embedding $v_i$. Other anchors, positive or not, are designated by $v_a$. Finally, $\Theta$ is a temperature hyper-parameter when computing the cosine similarity between two embeddings. We choose $\Theta=0.3$ since the training data is small and noisy, which avoids sharp features in the cosine separations of embeddings. This hyperparameter choice is mostly empirical. Additionally, batch size is set to $512$ and a learning rate of $10^{-3}$ is adopted, with a linear warm-up over 3 epochs for \zlbg.

The above expression \ref{eq:supcon_loss} of loss is further generalized to use a "relationships" matrix per batch of index $b$ with $N_b$ elements. The expression developed in this work is akin to the formulation described by \cite{srinivasa2023_cwcl} as Continuously Weighted Contrastive Loss (CWCL). We define $(r_{ij})\in \mathcal{M}_{N_b}([0,\,1])$ relationship weights that describe the label similarity of our elements. Instead of "hard" clustering ($r_{ij}\in\{0,\,1\}$) of positives and negatives in the dataset, we employ "soft" clustering ($r_{ij}\in[0,\,1]$). Similarity between objects is now a continuous mapping between 0 (pure negatives) and 1 (pure positives) rather than a strict boolean, which helps shape the projected space. Therefore, the loss function can be written as:
\begin{equation}\label{eq:wscl_loss} 
\mathcal{L}
=
- \frac{1}{|I|}
\sum_{i \in I}
\sum_{j\in A(i)}
\frac{r_{ij}}{\sum_{k \in A(i)} r_{ik}}
\left[
\log \left(\frac{\exp \left(v_i \cdot v_j / \Theta\right)}{\sum_{a \in A(i)}
\exp\!\left(
v_i \cdot v_a / \Theta
\right)} \right)
\right]
\end{equation}
Concretely, the above expression allows to define smoother relationships between elements, and obtain a continuous representation of positives and negatives. The sample is labelized in two ways: by type and by redshift, which builds the relationship coefficients $r_{ij}$. We discuss weighting in the following subsection.

\subsection{Weights}\label{sec:architecture:weights}

With weighted supervised contrastive learning, the embeddings can be aligned to represent the data-space in a meaningful way using continuous weights instead of hard weights. Since the primary objective is type classification and redshift performance, we introduce weighting schemes for both in this section.

\paragraph{Class weighting} The 5 labels used to labelize galaxy types are "ELG", "QSO" (contaminants) and "LBGa", "LBGse" and "LBGme", characterizing different absorption / emission strengths in LBG as described in \ref{sec:motivation}. The set of class labels is noted $\mathcal{Y}=\{$LBGa, LBGme, LBGse, ELG, QSO$\}$. Contaminants are considered as pure negatives (null weight): with this, ELGs and QSOs are well separated with the rest of the embedding space. Moreover, there is no redshift weighting applied to the ELG/QSO populations. For ELGs, most ELG contaminants are found at low redshift ($z_\mathrm{ELG}\sim0-0.2$) due to confusing [\ion{O}{2}] \wl{3726,3729} with \lya \wl{1216} that falls in the blue detector (at $z_\mathrm{LBG}\sim2$). For QSOs, they are at varying redshifts, but one wants to focus the shape of the spectra rather than the sole redshift inference, so they are kept in the same bin. However, the LBG population is intrinsically varied and considering all LBGs to be of the same type makes for harder generalization of embeddings, so discriminating between different classes (LBGa, LBGme, LBGse) while still allowing some self-similarity (compared to how contaminants are dealt with) helps the overall architecture. Therefore, the network should know that LBG subclasses are different, but LBGme-LBGse is less different than LBGme-QSO, for example. We define $\mathbf{C}=(c_{mn})_{(m,n)\in \mathcal{Y}^2}$, a class similarity matrix between types of galaxies. Here, $m,n$ are galaxy type labels. For all available labels $m$, $c_{mm}=1$: two galaxies of the same type should not be distinguished in terms of class-driven weights. The following coefficients are chosen empirically:

\begin{equation}\label{eq:fid_rel_matrix}
\begin{array}{c|ccccc} 
 & \mathrm{ELG}  
 & \mathrm{QSO} 
 & \mathrm{LBGa} 
 & \mathrm{LBGme} 
 & \mathrm{LBGse} \\
\hline
\mathrm{ELG}   & 1 & 0 & 0 & 0 & 0 \\
\mathrm{QSO}   & 0 & 1 & 0 & 0 & 0 \\
\mathrm{LBGa}   & 0 & 0 & 1 & 0.5 & 0.5 \\
\mathrm{LBGme} & 0 & 0 & 0.5 & 1 & 0.5 \\
\mathrm{LBGse}   & 0 & 0 & 0.5 & 0.5 & 1
\end{array}
\end{equation}

This choice is motivated by visual inspection of the uniform manifold approximation and projection (UMAP \cite{mcinnes2020umapuniformmanifoldapproximation}) figure of the embeddings when training without $\mathbf{C}$-weights. The LBG-type galaxies have significant overlap in embedding space, so imposing strict separation ($\mathbf{C}=\mathbb{I}_5$) is not natural for training. This constraint is alleviated with weighting. The latter can be further fine-tuned, but such manipulations can easily become biased by performance over a single, small test dataset and not generalized to larger datasets. In order to avoid fine-tuning, we choose a simple matrix configuration that performs well (and significantly better than without weighting, especially for redshift performance) and that does not cater to a particular LBG sub-type. It is noted that other configurations of weighting for $\mathbf{C}$ can yield similar or better performance: further discussions about this choice can be found in Appendix~\ref{sec:appendix:class_weighting}. Thus, we restrict our choice to a simple matrix in order to not over-tune on the test dataset.

\paragraph{Redshift weighting} Labelization of the data also comes from redshift. To preserve redshift information within the LBG projections, a separate redshift head with a secondary loss function for joint learning was considered. This approach ultimately proved less efficient than our selected strategy: directly incorporating the redshift information into the relationships. 
For an anchor galaxy at redshift $z_a$ and a galaxy at redshift $z_i$, we define the redshift relationship weight as
\begin{equation}
w(z_a,z_i)=
\begin{cases}
\displaystyle \exp\left[-\frac{1}{2\sigma_z^2}{\left(\dfrac{z_a-z_i}{1+\bar{z}}\right)^2}\right] & \text{Gaussian kernel} \\
\displaystyle \mathbbm{1}\left(\left|\frac{z_a-z_i}{1+\bar{z}}\right|\leq \Delta z\right) & \text{Top-hat kernel}
\end{cases}
\end{equation}
where $\bar{z}=(z_a+z_i)/2$ is the mean redshift of the galaxy pair. Here, $\sigma_z$ or $\Delta z$ are chosen as arbitrary hyperparameters that match with the chosen "refinement" prior $\lz=0.025(1+z_\mathrm{pred})$ size for \redrock ($\mathcal{U}(z_\mathrm{pred}-\lz, z_\mathrm{pred}+\lz)$). The amplitude pre-factor of $\lz$ is chosen to match $\lz=0.1$ at redshift $z_\mathrm{pred}=3$, used in \lbgnet \cite{RuhlmannKleider2024_LBGs}. The latter chose a fixed half-width ($\lz=0.1$) instead of a redshift dependent width. Further discussion on kernel choice and size is presented in Appendix~\ref{sec:appendix:redshift_weighting}, while the prior half-width amplitude $\lz$ is investigated in Appendix~\ref{sec:appendix:lz}. The fiducial choice presented in Section~\ref{sec:results} is the gaussian kernel with $\sigma_z=0.025$, matching well with the amplitude of $\lz$.

To summarize, if $i,j$ are two galaxies in the training sample, the relationship coefficient between the two are defined as:
\begin{equation}\label{eq:rel_weights}
r_{ij} =
\mathbbm{1}({i\ne j})\, c_{y_i y_j}
\begin{cases}
w(z_i,z_j) & \text{if } (y_i, y_j )\in \mathbf{G} \\
1 & \text{otherwise}
\end{cases}
\end{equation}
\noindent where $\mathbf{G}=\{ \mathrm{LBGse,\,LBGme,\,LBGa}\}$ is the "good" galaxies. Additionally, $\mathbf{B}=\{ \mathrm{ELG,\,QSO}\}$ are defined as the contaminants ("bad"). Here, $w$ is the redshift kernel, $i,j$ are the galaxy indices at redshifts $z_i, z_j$ and $y_i,y_j\in\mathcal{Y}$ the associated galaxy type labels. The indicator function $\mathbbm{1}$ verifies we do not include self pairs in the relationships (diagonal is null), since they carry no information when building the contrast over the embeddings.

\subsection{Classification}\label{sec:architecture:classification}

The classification step is a downstream task, after the $\mathbf{Rep}$ layer is trained. It is done through a multi-layer perceptron (MLP). Each spectrum representation $h=\mathbf{Rep}(x)$ is mapped to a 5-dimensional probability vector $p(x)$ where each of the components corresponds to a target class (ELG, QSO, LBGse, LBGme, LBGa). This is an improvement on the original classification, since \lbgnet provides a single valued confidence level and does not give information about the galaxy type. This allows to both characterize our contaminant population, and provide an estimate of the LBG target class. The MLP itself is small, containing two hidden layers of dimensions 128, 64 units with GELU activation \cite{hendrycks2023gaussianerrorlinearunits} and a \texttt{Dropout} after each layer at rate $0.3$. This is then projected to the number of classes with a \textit{softmax} activation function.
While the MLP performs well on contaminants, it performs poorly at identifying the correct subtype of LBG. Given the LBG types are loosely defined in the VI sample, and the purpose is primarily to discriminate contaminants (and not LBG subtypes, although useful), the MLP offers sufficient classification performance. Single galaxy type classification performance is presented in Section~\ref{sec:results:classification}.

Alternative classification schemes are also explored. For example, one can use cosine similarity over the embeddings and directly compare to the spectra of the training dataset. While this method directly leverages the architecture of the embeddings, it is ultimately not a generalized approach and is therefore limited by the training dataset examples. Thus, we prioritize the MLP network for classification.

To compare directly with the confidence level of \lbgnet ($\tau_\mathtt{lbgNET}$), we will further combine the $p$ vectors into a single $s_\mathtt{zlbg}(x)$ classification confidence score. Once normalized such that $\sum_{n\in\mathcal{Y}} p_{x,\,n}=1$, the confidence levels are merged as:
\begin{equation}\label{eq:s_zlbg}
    s_\mathtt{zlbg}(x)=\frac{\sum_{n\in\mathbf{G}} p_n(x)}{\sum_{n\in\mathbf{G}} p_n(x)+\sum_{n\in\mathbf{B}}p_n(x)}\in[0,1]
\end{equation}
Other classification metrics are considered, but similar performance is generally observed\footnote{Another well performing considered metric is $s^{\mathrm{max}}_\mathtt{zlbg}(x)=\max\limits_{n\in\mathbf{G}}(p_n(x))\left(1-\max\limits_{n\in\mathbf{B}}(p_n(x))\right)$.}.
Such a metric is great for classification, but does not take into account redshift performance: we tackle this issue in the following Section~\ref{sec:architecture:z_inference}. On the contrary, the $\tau_\mathtt{lbgNET}$ is intrinsically linked to redshift performance and classification, being the confidence level of an existing emission or absorption line. 

 
\subsection{Redshift inference}\label{sec:architecture:z_inference}

Redshift predictions of the spectra are obtained with a K-Nearest Neighbors (KNN)\footnote{Other architectures have been considered (such as \texttt{XGBoost}\cite{Chen_2016_XGBoost} or MLPs) and found to be less effective than the KNN.} network that we name z-KNN in Figure~\ref{fig:architecture}. Because we enforce the embeddings to learn redshift proximity through the weights described in Section~\ref{sec:architecture:weights}, a nearest neighbor architecture on the embedding space is naturally well suited to learn a redshift representation for spectra. The KNN directly uses cosine distance, just like contrastive loss, for the neighbor distance metric. However, CL computes the cosine distances on the $\mathbf{Proj}$ layer, whereas cosine distance for the KNN is computed on the $\mathbf{Rep}$ layer. By default, the KNN uses 50 neighbors.

For now, this does not include any information about redshift confidence, since the currently defined threshold is the classification score $s_\zlbg$. While getting accurate error quantification with the KNN is not easy, developing a proxy for redshift error is important to assess our confidence in the inferred redshift. To do so, for each new spectra $x$ we obtain $\sigma_\mathrm{KNN}(x)$: the redshift neighborhood scatter. The KNN maps newly ingested spectra $x$ (e.g. from the test dataset) to a neighborhood in the inferred redshifts provided by the fitted sample (training set representations from $\mathbf{Rep}$). All of these neighbors are at a defined distance $d_k$ and $k\in\mathbf{K}(x)$ is the set of 50 neighbors from the fitted sample to $x$. After normalizing $d_k$ such that the sum over $\mathbf{K}(x)$ is unity and defining $\omega_k=1/d_k$, $\sigma_\mathrm{KNN}(x)$ can be written as:
\begin{equation}\label{eq:sigma_knn}
    \sigma_\mathrm{KNN}(x)=\sqrt{\sum\limits_{k\in\mathbf{K}(x)}\omega_k(z_k-\mu)^2} \text{ with } \mu=\sum\limits_{k\in\mathbf{K}(x)}\omega_kz_k
\end{equation}
This scatter quantifies how uncertain the redshift evaluation is: if a spectra representation is very close in $\mathbf{Rep}$-space ($\omega_k$ is large) but at a very different redshift from the mean surrounding redshift $\mu$, $\sigma_\mathrm{KNN}(x)$ will increase. To quantify this into thresholds for selection, we develop a redshift quality value $q_z\in[0,1]$ defined as:
\begin{equation}\label{eq:qz_flag}
    q_z(x)=\exp\left(-\frac{\sigma_\mathrm{KNN}^2(x)}{2\lz^2}\right)
\end{equation}
The redshift quality is here dependent on the prior threshold $\lz$ half-width selected for the analysis. The redshift selection threshold is then defined as:
\begin{equation}\label{eq:tau_zlbg}
    \tau_\zlbg(x)=s_\zlbg(x)q_z(x)
\end{equation}
For \zlbg, this means that two selections are applied: first the confidence that the observed spectra is an LBG (classification step, over $s_\zlbg$), then in second that we can recover a good redshift (over $\tau_\zlbg$).
\section{Data}\label{sec:data}

In this section, we present the data used for this analysis, as well as the augmentations performed on said data. The templates used are also described. As previously mentioned, we focus on \setzero (complete visual inspection of a pilot program), \setone (partial inspection of some pilot programs) and \settwo (survey validation ELGs and QSOs). 

Contrastive learning benefits from data augmentations, as it leverages augmented views of spectra to become insensitive to certain features in the embedding space. However, the augmentations must be representative of the spectra used. We use templates to augment the spectra. In the following Section~\ref{sec:data:templates}, we describe the templates used for this analysis. In Section~\ref{sec:data:training_quality}, quality cuts and masks to the dataset are presented, in order to create the stem of the training and validation sets. Section~\ref{sec:data:augmentations} uses this cleaned set of spectra to build the data augmentations in redshift and SNR. The data presentation is concluded with  Section~\ref{sec:data:training_testing} showcasing the fiducial sample split into test, training and validation datasets.

\subsection{Templates}\label{sec:data:templates}

The templates presented in Section~\ref{sec:motivation} (Figure~\ref{fig:motivation:zoom_features}) are four LBG templates built on the visually inspected samples: 1 is dedicated to LBGa, two to LBGme and 1 to LBGse. The templates are built using an iterative process. First, a dataset of visually inspected spectra are stacked into 4 templates at rest frame, binning by measured emission width (EW) of the $\lya$ feature. The emission width is measured using an asymmetric Gaussian profile fit to the $\lya$ emission. Further measurement details are described in Section 4.4 of \cite{RuhlmannKleider2024_LBGs}. These EW bins are as follows: No $\lya$ emission, EW $\in[0.0, 18.0]$, EW $\in[18.0, 40.0]$ and EW $\in[40.0, 250.0]$. In particular, Section 4.2 of \cite{RuhlmannKleider2024_LBGs} describes the process of obtaining the first version of these templates. Secondly, we use the now trained version of \lbgnet developed in \cite{RuhlmannKleider2024_LBGs} and apply the network to a sample of spectra yet unclassified by visual inspection. A stringent $\tau=0.99$ threshold is chosen to only keep high quality classifications of the spectra. The new spectra obtained are again binned by EW and included in the original stacks, thus further refining the templates built. A total of 5558 spectra are used, out of which 1123 have no $\lya$ emission (LBGa) and the remaining 4435 are approximately equally distributed among LBGme1, LBGme2 and LBGse. The procedure is extensively described in Section 4.4 of \cite{RuhlmannKleider2024_LBGs}. The templates obtained are both useful for redshift augmentations of the training and validation sets, as detailed in the following subsections, but also as the templates used by \redrock during the fitting procedure after the prior has been obtained, since these are representative LBG templates. In contrast to \cite{RuhlmannKleider2024_LBGs} which used the Principal Component Analysis (PCA) space parameterization of similar templates as shown in Figure 13, we use the templates in physical space and do not reproject to 4 components in PCA space. Both can be used by \redrock.

\subsection{Training Quality}\label{sec:data:training_quality}

Before any augmentations, we discuss pre-processing steps to improve the quality of the training set. The augmentations will then be derived from the set of cleaned spectra. The goal of this section is to remove spurious features such as detector noise, spectroscopic sky lines or missing pixels.

\paragraph{Spectral binning} First, just like \cite{busca2018_quasarnet}, we re-bin the spectra using its inverse variance (ivar) to weight the co-addition of flux pixels. We use a linear coarse binning of $\mathrm{d}\lambda=8\angstrom$. Spectra in DESI have a wavelength binning of $\sim0.8\angstrom$ \cite{Spectro.Pipeline.Guy.2023}. The original \qnet \cite{busca2018_quasarnet} algorithm used logarithmic binning to bin the spectra down to 443 pixels from $\sim4400$ pixels, applying a similar reduction factor before ingestion. The logarithmic binning is historically due to the structure of the BOSS spectra \cite{BOSS.GilMartin.2020} used for training, but here we follow the \lbgnet structure for input spectra and \qnet's current implementation in DESI, where linear binning is used. Finer binning ($\mathrm{d}\lambda=2\angstrom$) was considered: target selection confuses the [\ion{O}{2} \wl{3726,3729}] doublet with $\lya \wl{1216}$, so in theory finer binning would have enough resolution to distinguish the [\ion{O}{2}] lines. In practice, we observe that performance is worse in both redshift and classification (augmentations are degraded, information is lost in noise) for both pipelines, hence we conserve the choice of $\mathrm{d}\lambda=8\angstrom$, which is a similar binning ratio to \qnet.

\paragraph{Spectrum ingestion} Following \lbgnet and \qnet, rebinned spectra are normalized by subtracting the weighted mean of the spectra and then dividing the resultant spectra by its weighted root mean square. The rebinned inverse variance (ivar) is used for weighting. The spectra shown throughout this work have this pre-processing applied, so flux is in arbitrary units and continuum is not necessarily above 0. This also makes for easier continuum connection with templates during augmentation of the spectra. 

\paragraph{Spectrum quality} Any spectrum with over 10 reported dead pixels after binning (ivar is null) is removed from the training set. Moreover, for each spectrum visually inspected, a quality score is attributed by the reviewer ranging 0 to 4. We select spectra where the mean quality $q_\mathrm{VI}$ over the reviewers surpasses 2.5 for all VI datasets included in this analysis (\setone, \setzero, \settwo), following \cite{RuhlmannKleider2024_LBGs}. This biases the analysis to spectra that were observed where the quality was sufficient to ascertain a good redshift measurement. However, there is contamination from bad spectra, noisy detections, low SNR... These are mainly labeled as "BAD" in the VI datasets. We investigate their impact on purity and efficiency in appendix~\ref{sec:appendix:bad_spectra}.  

\paragraph{Spectrum masking} Finally, some parts of the spectrum are masked due to sky or detector features. The masking procedure replaces the selected pixel region with gaussian noise over the underlying interpolated continuum between the pixels outside the mask. The gaussian noise level is derived from the smoothed ivar computed with a smoothing window of $100\angstrom$ over the real ivar, thus averaging over variance spikes. This masking is applied to the spectra in 3 different areas:
\begin{itemize}
    \item Camera overlaps: following methodology by \cite{RuhlmannKleider2024_LBGs}, a 40\angstrom window centered on the overlap of the camera bands is masked and replaced with gaussian noise, following the procedure described above with the smoothed ivar. This avoids spurious features due to detector edges.
    \item Inverse-variance spikes: when the ratio of the smoothed ivar over the underlying ivar is larger than 3\footnote{This is binning-dependent. For $\mathrm{d}\lambda=2\angstrom$ we choose a limiting ratio of 5.} at the same wavelength, we apply gaussian masking over a small window of width $2\mathrm{d}\lambda$. This detects spurious flux features due to sky lines and potential incorrect correction for the presence of sky lines in the data reduction. To avoid shifting the wavelengths of sky lines in the augmented spectra dataset, which could bias the network, we mask these regions. These regions are redshift-independent, which avoids biasing redshift measurements.
    \item Some spectra still report a few pixels with null ivar. We replace these pixels with gaussian noise based on the underlying smoothed ivar.
\end{itemize}
The above spectral masking is only applied to the validation and training datasets, which are both augmented in redshift and SNR. Such masking is not applied to the test dataset. 

\subsection{Augmentations}\label{sec:data:augmentations}

\paragraph{SNR} One good example of this is the signal-to-noise (SNR) ratio, where same spectra at different noise levels are counted as a fully positive pair. To do this, spectra are co-added through the DESI spectroscopic pipeline \cite{Spectro.Pipeline.Guy.2023} with varying number of exposures. In order to augment spectra in SNR, co-adds give multiple equivalent $60$ min exposures, $120$ min and $300$ min exposure by dividing the total exposure time in smaller blocks. 

\paragraph{Redshift} Redshift augmentations are also included in the training dataset. This augmentation has a dual purpose of both augmenting the number of dissimilar spectra by generating new spectra at different redshifts (as presented in Section~\ref{sec:architecture:weights}) and preparing the network to learn spectra representations over the whole expected redshift, despite the limited redshift coverage for $z\gtrsim3.8$ in the current VI samples. In contrast to SNR augmentations, redshift augmentations are only performed on the LBG-type spectra.

For each spectrum at redshift $z_\mathrm{spec}$ of the training and validation sets, a total of $n_\mathrm{aug}=5$ spectra including the original spectrum are made. In order to keep newly made spectra close to the original, the $4$ augmented redshifts are drawn from a uniform distribution around $z_\mathrm{spec}$, in  $\mathcal{U}(z_\mathrm{spec}-0.8,\,z_\mathrm{spec}+0.8)$, and bounded by $z\in[2,4.5]$. The $0.8$ range is arbitrarily chosen to reach the higher ends of the redshift range, which has low LBG coverage. Redshifts are augmented up to 4.5 to include training for g-dropout selections which go to further redshifts. Currently, VI datasets mostly include u-dropout based target catalogs, which explains the lower coverage at high redshifts. The validation and training set are both augmented, but the test set remains untouched. 


Evidently, redshift sampling is an issue for spectral coverage, since at the new redshift these spectra do not cover the complete wavelength range anymore (3600-9800$\angstrom$). Thus, augmented spectra need to be completed by templates to match the spectral coverage. The LBG stacked spectra presented earlier in  Figure~\ref{fig:motivation:zoom_features} and Section~\ref{sec:data:templates} match well the provided spectra. For each spectra in the test dataset, the reviewer visually assigned the "LBGa, LBGme or LBGse" tag to the spectrum. One of the templates is then chosen according to the tag of each spectrum to extend over the missing redshift coverage\footnote{In the case of LBGme, two templates were originally made during stacking (as shown in Figure~\ref{fig:motivation:zoom_features}). We select one randomly during augmentation for diversity.}. Over a blending window of width $400\angstrom$, the template is rescaled to match the amplitude of the now augmented spectrum. In this window, the template and the new spectrum are merged using a cosine blend function\footnote{A cosine blend between two functions $f_1$ and $f_2$ merges these functions into $f(t)=(1-\mu(t))f_1(t)+\mu(t)f_2(t)$ where $\mu=(1+\mathrm{cos}(\pi t))/2$. Here, $t\in[0,1]$ linearly evolves over wavelength on the blend window.}. The connection between the template and the original spectra is seamless, with smooth edges to the blending window. Thus, the template does not learn connection artifacts. For variations in the templates and noise, we refer to the original inverse variance of the spectrum to add synthetic gaussian noise to the template such that the noise level matches the reported per-pixel inverse variance. Over the blending window between the two spectra, the noise level takes into account contribution from the original spectrum and adds less noise to the merged template.

A caveat to this process is augmenting from a lower to a higher redshift. In that case, the original spectrum is shifted to higher wavelengths, therefore the connection happens on the lower edge of the wavelength grid. That part of the grid is the blue detector in DESI \cite{DESI2016b.Instr, Spectro.Pipeline.Guy.2023}. In that area, the inverse variance sharply drops and the spectrum is very noisy \cite{DESI2022.KP1.Instr}. Hence, before augmentations from a low redshift to a high redshift, the low end of wavelength coverage (3600-4000$\angstrom$) is cut off and the blending happens around 4000$\angstrom$.

Finally, we verify the good representation of the augmented training spectra by performing a Principle Component Analysis (PCA) of both the augmented and original spectrum. Projecting to the first two components of PCA, we verify good overlap of the two datasets and no significant dependency in redshift of the augmented spectra.

\subsection{Training and testing}\label{sec:data:training_testing}

\begin{figure}[h]
    \centering
    \includegraphics[width=0.98\textwidth]{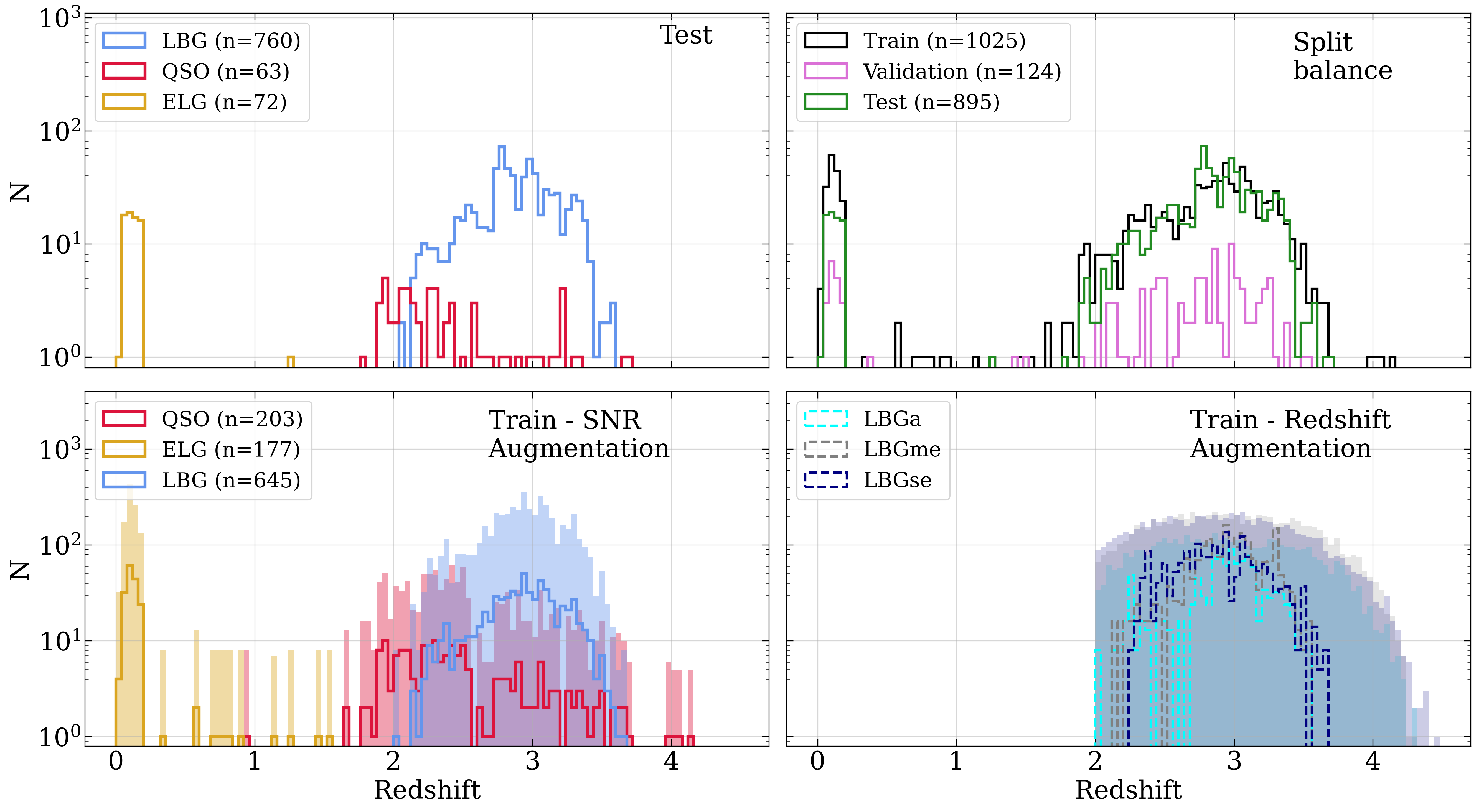}
    \caption{\textit{Top:} (\textit{Left}) The test dataset distributions, with no augmentations. $n$ represents the number of unique sources. (\textit{Right}) Split balance of unique spectra between test/train/validation. \textit{Bottom:} (\textit{Left}) The training set, with the LBG and contaminants. (\textit{Right}) Showcasing the redshift augmentations over the LBG type spectra. Here, $n_{\mathrm{aug}}=5$ is the ratio between the number of redshift augmentations and SNR augmentations.}
    \label{fig:data:hist}
\end{figure}

In this subsection, we cover the fiducial train/validation/test split. We display the redshift distributions of the different galaxy populations included in the datasets used in Figure~\ref{fig:data:hist}.

\paragraph{Test set} Out of the test dataset, 548 LBGs come from \setone. An additional 212 LBGs come from \setzero (remaining from the dataset split on the original dataset). The contaminants come either from the dedicated DESI survey validation tiles for ELGs and QSOs (explored in \cite{VIQSO.Alexander.2023, VIGalaxies.Lan.2023}, presented as \settwo in Section~\ref{sec:motivation}), or alternatively from the ELGs and QSOs that were not used in training from \setzero and were reserved for additional testing. The contaminants injected from \settwo and \setzero are then resampled to match the densities of contaminants of the original VI tile (\setzero) (approximately $\sim7\%$ (63) QSOs and $\sim8\%$ (72) ELGs). Thus, we match expected contamination rates after target selection. As mentioned in Section~\ref{sec:motivation}, a full description of the origin of spectra for each dataset is given on \href{https://doi.org/10.5281/zenodo.21775480}{\textcolor{blue}{\texttt{zenodo}}}, as the training and test sets come from patchworks of multiple datasets. 

\paragraph{Training set} The training set used for \lbgnet in the case we display is not equivalent to the dataset that was used in \cite{RuhlmannKleider2024_LBGs}, since the fiducial network developed there incorporates all of the LBGs that were VIed in the training set, after two-fold cross validation of individual dataset splits. Therefore, we re-train \lbgnet in order to compare with \zlbg accurately.  Moreover, the training data for the original version of \lbgnet included some simulated LBG spectra as well as more augmentations. We decide to preserve the same training set for both pipelines to maintain a fair comparison, and therefore re-train \lbgnet. 

\paragraph{Validation set} A validation dataset is kept as a slice of the training set for \texttt{EarlyStopping} (with a patience of 10 epochs) of the pipelines, once the validation loss reaches a minimum\footnote{For the classifier head, the validation accuracy is monitored instead of the loss.}. 

The fiducial design described above of the training set is named "Base". Other configurations for training are also considered, detailed in the following list:
\begin{itemize}
    \item Base+Val : The validation and training data are merged into a single training set. There is no validation or early stopping. We use a maximum of 30 epochs for the \lbgnet encoder, \zlbg encoder and classifier.
    \item Base (no z-aug): We do not include any redshift augmentation of the LBGs, only augmentations in SNR.
    \item Base+Val (no z-aug): Combination of the two cited scenarios. Larger diversity in the training dataset due to the inclusion of the validation data into training, with less augmentations.
\end{itemize}
The impact of these other configurations is discussed in the following Section~\ref{sec:results}. During training, we use a class-balanced sampler with 10 redshift bins for LBGs in order to ensure redshift representativity. During training of the classifier layer for \zlbg, the batches also up-sample lacking contaminants by inverse frequency, in order to have a class-balanced classifier training.
\section{Results}\label{sec:results}

In this section, we present the de-contamination and redshift measurement performance over the test dataset. The metrics used to assess performance are \textit{purity} and \textit{efficiency}, akin to \cite{RuhlmannKleider2024_LBGs}. Similarly to respectively precision and recall rate in machine learning, they are defined as:
\begin{equation}
    \pi_{\mathrm{net}}^\mathrm{LBG}(t)=\frac{N_\mathrm{net}^{\mathrm{LBG}}(t)}{N_\mathrm{net}(t)} \text{ and } \varepsilon_{\mathrm{net}}(t)=\frac{N_\mathrm{net}^{\mathrm{LBG}}(t)}{N_\mathrm{VI}^{\mathrm{LBG}}}
\end{equation}
Here, "net" designates either \lbgnet or \zlbg. "VI" corresponds to numbers derived from the visual inspection campaign acting as the truth catalog for this study. As such, purity $\pi_\mathrm{net}$ is the number of objects selected that are true LBGs ($N_\mathrm{net}^{\mathrm{LBG}}$) over the number of objects ($N_\mathrm{net}$) selected by the network at a given confidence threshold $t$. At the lowest confidence level, every object is selected, and therefore $N_\mathrm{net}$ corresponds to the number of objects in the test dataset. Efficiency $\varepsilon_\mathrm{net}$ is defined as the number of objects selected by the network that are true LBGs ($N_\mathrm{net}^{\mathrm{LBG}}$) over the total number of true LBGs ($N_\mathrm{VI}^{\mathrm{LBG}}$). The confidence level used $t$ for \zlbg depends on the task: classification uses $s_\zlbg$ while redshift performance will use $\tau_\zlbg$. For \lbgnet, it is always $\tau_\lbgnet$.

The same metrics are defined for redshift misidentification where the LBG subscript is replaced by $\lz$. In that case, purity is the ratio of true LBGs selected by the network ($t_\mathtt{net}(x)\geq t_\mathtt{net}$) that respect $|z_\mathrm{true}-z_\mathrm{pred}|\leq \lz=0.025(1+z_\mathrm{pred})$ over the number of objects selected by the network that are also true LBGs. Efficiency is the same numerator as purity, but now divided by the number of (true) LBGs in the VI. One can think of purity as the percentage of correct identification in the selected sample, and efficiency as the percentage of correct predictions over the total possible identifications. 

The results are presented in the following order. First, classification performance is discussed in Section~\ref{sec:results:classification}, both as a direct comparison between the two pipelines and also as individual classification for \zlbg. Then, redshift performance is showcased in Section~\ref{sec:results:redshift}. In Section~\ref{sec:results:training_sample}, we investigate sensitivity to different training set configurations. Finally, Section~\ref{sec:results:joint} presents opportunities for joint information inference from the two pipelines. Presented results are all under controlled randomness in order to assess hyper-parameter effects, thus all runs are performed with the same random seed unless otherwise specified.

\subsection{Classification}\label{sec:results:classification}

\begin{figure}[h]
    \centering
    \includegraphics[width=0.98\textwidth]{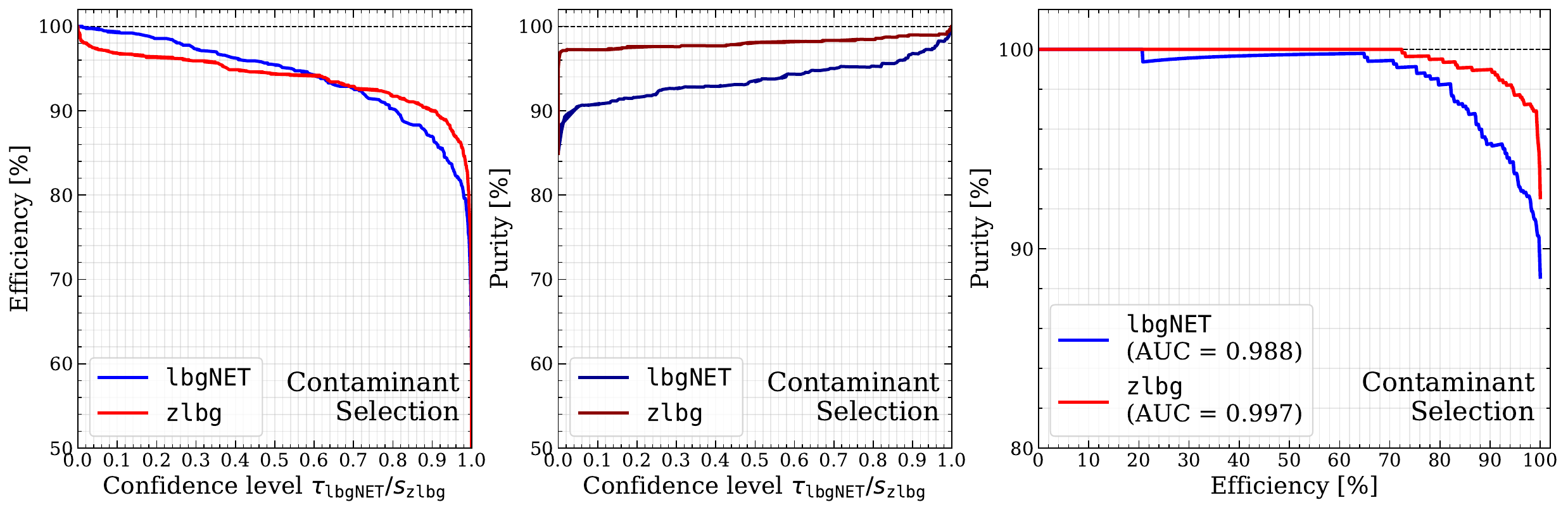}
    \caption{Efficiency and purity as a function of the confidence level (respectively left, center) and purity as a function of efficiency (right). The failure criterion is misidentifying an LBG for a QSO or ELG. The threshold used are $\tau_\lbgnet$ and $s_\zlbg$.}
    \label{fig:results:pur_eff_classification}
\end{figure}

Figure~\ref{fig:results:pur_eff_classification} reports the nominal purity / efficiency curves for the "Base" training set. In this section, selection is evaluated with $s_\zlbg$. It is also possible to use $\tau_\zlbg$: in that case, the performance change in classification is minimal (AUC=0.995).
\zlbg presents better overall classification of outliers than the \lbgnet pipeline. The AUC (Area Under Curve) score of the purity / efficiency curves is also computed (with $(\mathrm{AUC},\pi,\varepsilon)\,\in[0,1]$). A larger AUC is a better performing classifier. For contaminant selection, a random classifier would exactly give the ratio of LBGs over the full test dataset. In this case, target selection is assumed to have a $\sim85\%$ purity, per reported by visual inspection of data. Therefore, a random classifier would have $\mathrm{AUC}\simeq0.85$, which is also why the curve for purity/efficiency returns a purity of $\sim0.85$ at maximal efficiency. We perform runs with the same parameters at different training seeds and conclude the contaminant selection AUC varies at the $\sim0.2\%$ level.

Since the classifier outputs the $p_n(x)$ vectors, we test the raw classification without condensing $p_n(x)$ down to a $s_\mathtt{zlbg}$ metric, as described in Section~\ref{sec:architecture:classification}. The test dataset used in this analysis is combined from different pilot program origins: therefore, only spectra from \setzero (remainder of the training/test/validation split in "test") are sub-classed by LBG type in LBGs. That subset of 212 spectra is therefore able to test classification for LBG subtypes. In \setone, all LBG targets do not have further sub-type labels due to differing VI procedure, which is why this dataset was not used for training.

\begin{figure}[h]
    \centering
    \includegraphics[width=0.85\linewidth]{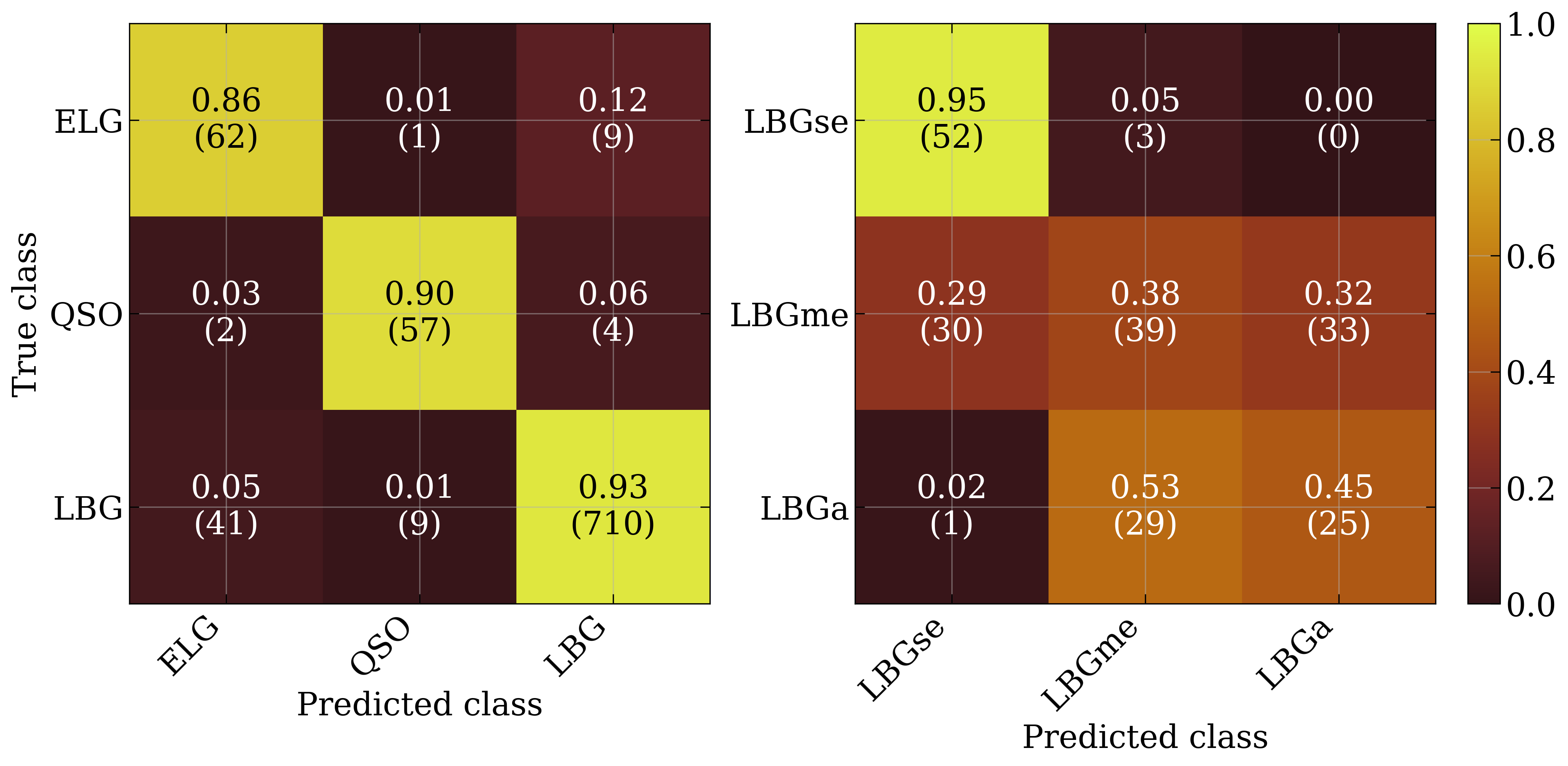}
    \caption{Reported confusion matrices on the test dataset. Numbers reported are the fraction ($\in[0, 1]$) of one type identified as a predicted type, and the number of single spectra associated to that percentage (in parenthesis). A perfect classifier should report a confusion matrix equal to the identity matrix $\mathbb{I}_3$. \textit{Left}: Classification over the full dataset, assuming classification as an LBG if $\arg\max\limits_{n\in\mathcal{Y}}p_n\in\mathbf{G}$ where $\mathbf{G}=\{\mathrm{LBGa,\,LBGme,\,LBGse\}}.$ \textit{Right}: Classification over the subset of LBGs in the test dataset that also have a subtype ($\mathbf{G}$) label. In this figure, the default configuration of the network is shown.}
    \label{fig:confusion_matrix}
\end{figure}

The rest of the test dataset can be used to test the $p_n$ vectors over $\{\mathrm{ELG, QSO, LBG}\}$ where $\mathrm{LBG}$ is any LBG sub-type. Here, since we showcase classification per galaxy type, we do not use the condensed threshold $s_\zlbg$ and keep the vector output from the MLP. Classification results by type of galaxy are presented in confusion matrices as displayed in Figure~\ref{fig:confusion_matrix}. The takeaway from these confusion matrices is that the network performs well when identifying QSOs and ELGs, but has difficulties with identifying LBG subtypes, especially differences between nearby emission width bins (LBGme-LBGa, LBGse-LBGme). In particular, the LBGme class spreads over its nearby bins (LBGse, LBGa) for the predicted class, and the LBGa galaxies bleed into the LBGme predictions. However, far away bins (LBGse-LBGa) are well discriminated, and there are many good predictions for LBGse. Since the boundary between these classes is continuous and only visually defined, overlap is expected. The more useful information mostly comes from identification of QSOs and ELGs to assess the nature of contaminants in the processed sample. Nonetheless, classification of LBGse elements in useful especially for further comparison with the dedicated LAE programs of DESI Run 2.

\subsection{Redshift performance}\label{sec:results:redshift}

\begin{figure}[h]
    \centering
    \includegraphics[width=0.98\textwidth]{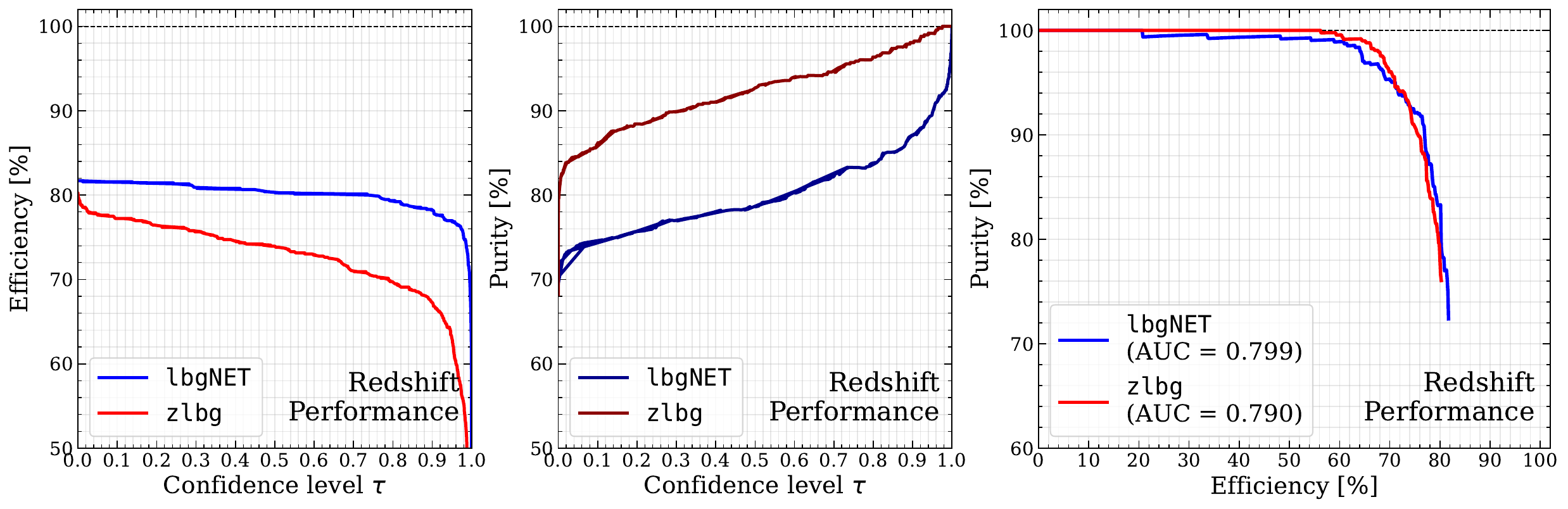}
    \caption{Comparison of purity and efficiency as a function of the redshift confidence level (left) and purity as a function of efficiency (right). The failure criterion is $|z_\mathrm{true}-z_\mathrm{pred}|>\lz$. The confidence levels used are $\tau_\lbgnet$ and $\tau_\zlbg=s_\zlbg q_z$.}
    \label{fig:results:pur_eff_redshift}
\end{figure}

After the $\mathbf{Rep}$ layer, a KNN regressor (architecture described in Section~\ref{sec:architecture:z_inference}) outputs the redshift based on the spectra representation. Figure~\ref{fig:results:pur_eff_redshift} presents the efficiency/purity curves for redshift performance, using the $\tau_\zlbg=s_\zlbg q_z$ threshold. Using solely $s_\zlbg$ as a confidence threshold significantly reduces redshift performance (AUC=0.779) hence justifying the $q_z$ quality flag to increase confidence on the redshift measurement.
The redshift efficiency does not reach 100\% due to the numerator definition of efficiency/purity for redshift performance, where in addition to the $|z_\mathrm{true}-z_\mathrm{pred}|\leq\lz$ criterion, the galaxies selected must be chosen by the network and be true LBGs. Because the purity/efficiency curve for redshift performance depends on the point estimate $z_\mathrm{pred}$ and the LBG selection, it is not obvious to define the performance for a random classifier, hence there is no baseline AUC to compare to. We perform runs with the same parameters at different training seeds and conclude the redshift performance AUC varies at the $2\%$ level. This is well explained by the dependency on classification performance as well as the general randomness of wrong predictions falling outside or inside the prior range.

\begin{figure}[h]
    \centering
    \includegraphics[width=0.98\textwidth]{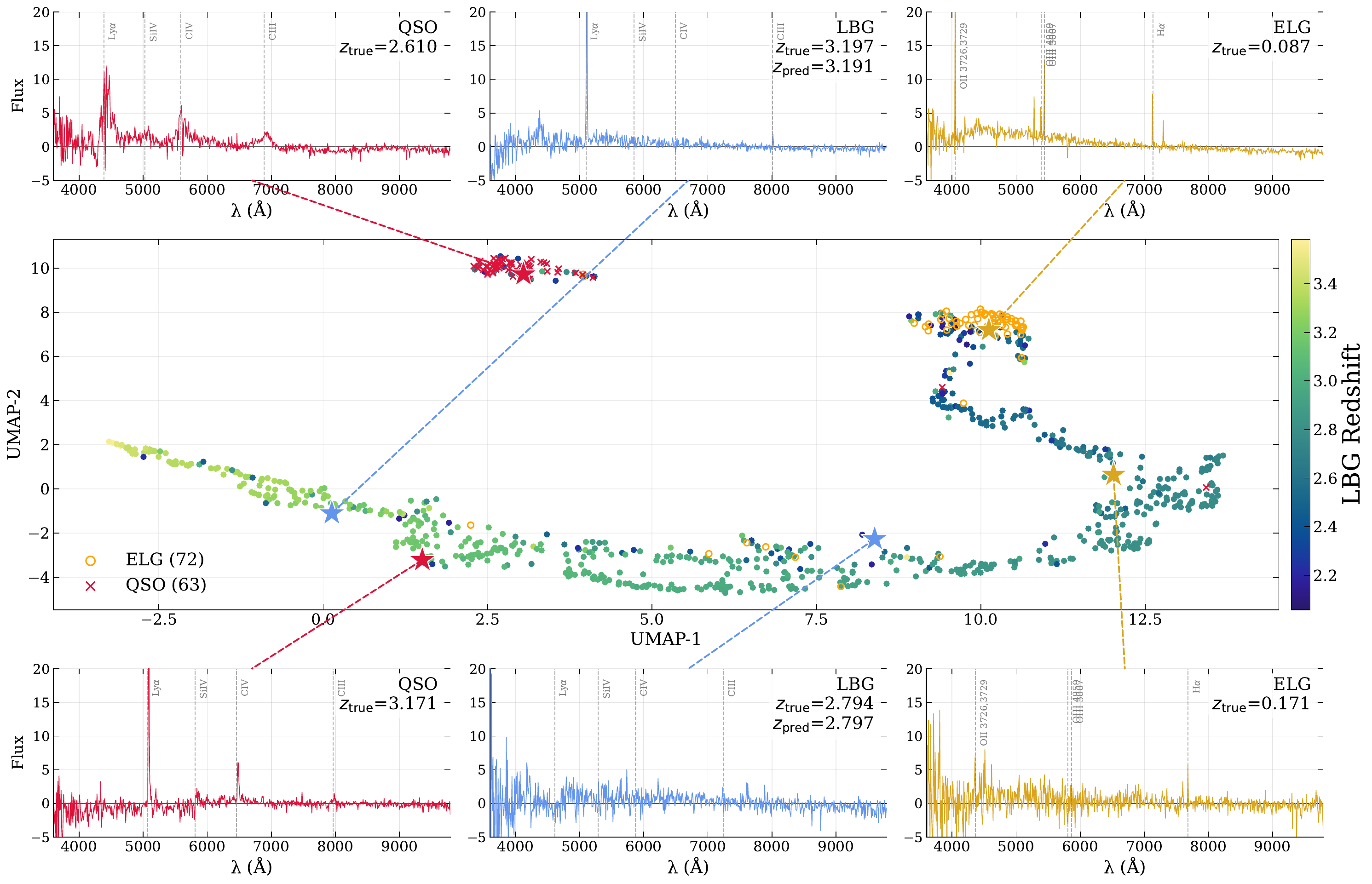}
    \caption{UMAP visualization of the test dataset's representations after the encoder ($\mathbf{Rep}$ layer). QSOs are highlighted as red crosses and ELGs as golden circles. The LBG population is colored by redshift. Example spectra, corresponding to their representations in the UMAP by star markers, highlight different success and failure modes and are discussed in text. The spectra are accompanied by some common spectroscopic line markers in observed frame at the true (VI) redshift.}
    \label{fig:results:umap}
\end{figure}

To further showcase the representation layer construction, a UMAP \cite{mcinnes2020umapuniformmanifoldapproximation} is used. The UMAP allows to condense the high-dimensional encoder vectors into a lower dimensional representation, here with two components, as shown in Figure~\ref{fig:results:umap}. In $\mathbf{Rep}$ space, it is clear that ELGs and QSOs are well separated from the LBGs. The crescent shape of the rest of the vectors represent the LBGs, color-coded by redshift. The representations carry the redshift gradient, enforced by the redshift kernel imposed in our weighting strategy detailed in Section~\ref{sec:architecture:weights}. To show some specific successes and failures in this representation space, a few example spectra are highlighted:
\begin{itemize}
    \item \textit{Top left:} The first example shown is a quasar at $z\sim2.61$. This is an example of an active galaxy nuclei with broad emission lines explained by the high velocity gas outflows around the supermassive central black hole. It is well discriminated and clusters with most of the other QSOs.
    \item \textit{Top center:} This spectra is an example of LBGse with a definite $\lya$ feature. The predicted redshift matches within the $\lz$ prior of the true redshift reported by the visual inspection.
    \item \textit{Top right:} A clear emission line galaxy contaminant within the ELG cluster of the UMAP. The strong [\ion{O}{2}] doublet explains the contamination in target selection, but the network is likely helped by good detection of $\mathrm{H}\alpha$ and the [\ion{O}{3}] doublet to distinguish the emission line galaxy.
    \item \textit{Bottom left:} This source classified as "QSO" in VI, is a narrow-line AGN with strong $\lya$, and is correctly classified by our method as an LBGse. Even during VI, such spectra are difficult to classify: the somewhat broad \ion{C}{3}] emission around $\sim8000\angstrom$ classifies this example as a QSO. Nonetheless, it falls in the LBG part of the UMAP.
    \item \textit{Bottom center:} An example spectra of an LBGa (no $\lya$ emission). The redshift is also correctly predicted.
    \item \textit{Bottom right:} An example of a failed ELG. The H$\alpha$ line is fairly faint, and the [\ion{O}{2}] doublet as well. There is no significant detection of the [\ion{O}{3}] doublet, which makes for a difficult spectra for the network.
\end{itemize}

The redshift gradient over LBGs presented in Figure~\ref{fig:results:umap} motivates further study: the representations are smeared over redshift, so changing the size of the $\lz$ prior width used for \redrock could recover more sources within the redshift range. On the contrary, \lbgnet is built around spectroscopic lines, therefore the network is apriori more susceptible to catastrophic outliers than \zlbg (for example, a line miss-identification). Variations of $\lz$ and \redrock performance based on the $\lz$ prior are investigated in Appendix~\ref{sec:appendix:lz}. 

\subsection{Sensitivity to training sample}\label{sec:results:training_sample}

\begin{figure}[h]
    \centering
    \includegraphics[width=0.98\textwidth]{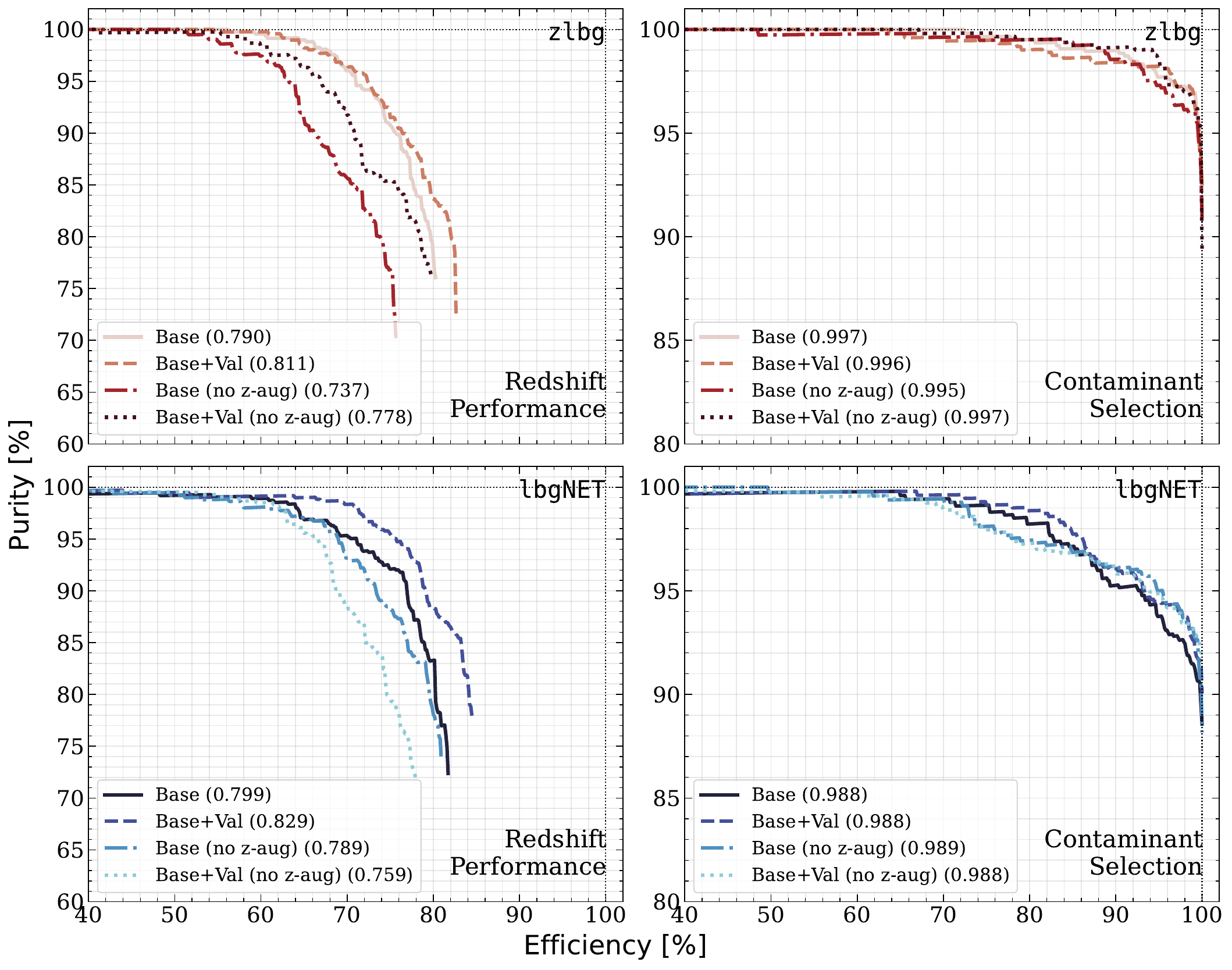}
    \caption{Performance evolution according to the dataset used. Base is the augmented, fiducial training set. Adding Val means moving the validation dataset into the training set and removing the \texttt{EarlyStopping} procedure. No z-aug signifies the augmentations in redshift presented in Section~\ref{sec:data:augmentations} are removed, and only SNR augmentations are preserved.}
    \label{fig:results:pur_eff_training_sets}
\end{figure}

Figure~\ref{fig:results:pur_eff_training_sets} presents the impact of different training set choices on redshift and contaminant efficiency / purity curves. \zlbg outperforms \lbgnet in contaminant selection, by design. This effect is especially visible for \lbgnet when comparing "Base+Val (no z-aug)" to "Base+Val", for example. The best performance on contaminant discrimination is reported when excluding redshift augmentations, for both \lbgnet and \zlbg. However, on the other side, redshift performance significantly improves when including redshift augmentations, especially for \zlbg. This is verified for both pipelines as well: both "Base" and "Base+Val" without z-aug under-perform compared to their counterpart. Training with larger datasets to assert these behaviors with more robustness would be useful. This is left for future work, when a larger sample of LBG galaxies has been collected by DESI (for example, during DESI Run 2 survey validation).

\subsection{Joint information}\label{sec:results:joint}
\begin{figure}[h]
    \centering
    \includegraphics[width=0.98\textwidth]{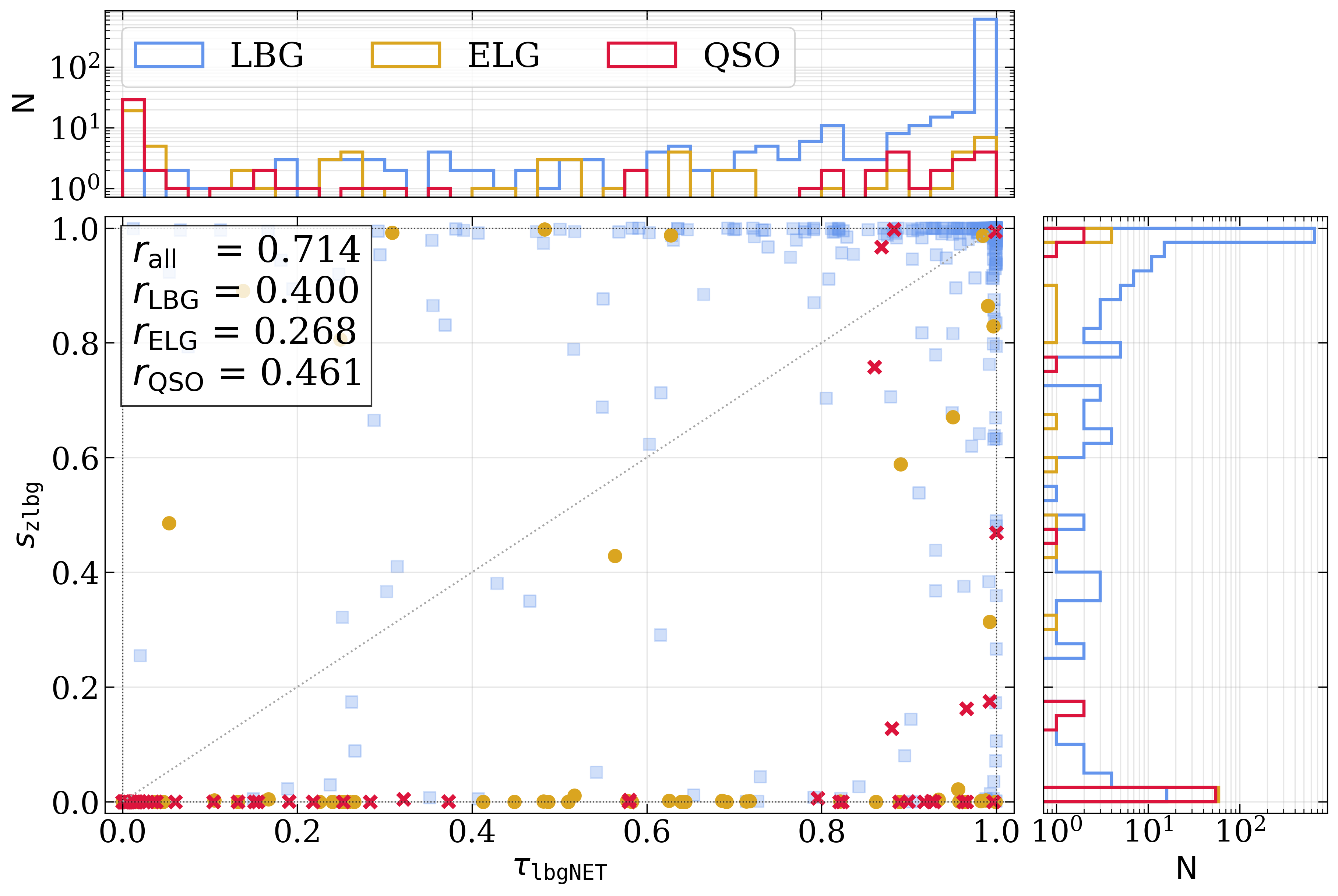}
    \caption{Confidence level of (in classification, $s_\zlbg$) \zlbg as a function of the confidence level of \lbgnet ($\tau_\lbgnet$). The contaminants are scattered in red crosses (QSO) and golden circles (ELG). The LBGs are represented by blue squares. The confidence levels are reported as marginal plots for both \lbgnet and \zlbg. For each (sub)sample, the Pearson correlation coefficient is given between the two methods.}
    \label{fig:results:confidence}
\end{figure}

In Figure~\ref{fig:results:confidence} we present the correlation between the confidence levels, in order to assess the difference in method extraction. We also report in the figure the Pearson correlation coefficient, defined as:
\begin{equation}
r_{\Gamma}
=
\frac{
\sum_{i\in\Gamma}
\left(
\tau^{(i)}_{\mathtt{lbgNET}}
-
\overline{\tau}_{\mathtt{lbgNET},\Gamma}
\right)
\left(
s^{(i)}_{\mathtt{zlbg}}
-
\overline{s}_{\mathtt{zlbg},\Gamma}
\right)
}{
\sqrt{
\sum_{i\in\Gamma}
\left(
\tau^{(i)}_{\mathtt{lbgNET}}
-
\overline{\tau}_{\mathtt{lbgNET},\Gamma}
\right)^2
}
\sqrt{
\sum_{i\in\Gamma}
\left(
s^{(i)}_{\mathtt{zlbg}}
-
\overline{s}_{\mathtt{zlbg},\Gamma}
\right)^2
}}\in[-1,1]
\end{equation}
\noindent where $\Gamma$ is the indices of the test dataset itself, or a subset of it, and $\overline{\tau}$ (or $\overline{s}$) is the average of confidence thresholds. For example, we compute $r_\mathrm{QSO}$: the correlation coefficient restricted to the VI quasars $\Gamma=\{\mathrm{QSO}\}$. We report the associated numbers on Figure~\ref{fig:results:confidence}. Marginal plots report the distribution of confidence levels according to galaxy type. While the joint information inference is left for future work, it is relevant to show that \lbgnet and \zlbg are complimentary approaches and do not leverage the same information: if they were, the respective Pearson correlation would be equal to 1. Nonetheless, the results are correlated (purely uncorrelated results would result in $r=0$), but many sources are correctly identified by only one of the two networks at a given threshold, despite the methods sharing the same backbone encoder. This is therefore solely an effect of the loss function. In particular, ELGs and QSOs are discriminated much more easily with \zlbg, as almost all contaminants are set to $\tau_\mathtt{zlbg}\sim0$ but spread over $[0,1]$ for $\tau_\mathtt{lbgNET}$. This also explains the low correlation values observed when restricting to the small QSO and ELG datasets.
\section{Conclusion}\label{sec:conclusion}

In preparation for DESI Run 2, neural networks will be crucial to identifying and estimating redshifts for faint high-redshift galaxies. The spectroscopic redshift pipeline \redrock \cite{Redrock.Bailey.2024} is more prone to systematics at these redshifts due to having a wider range of redshifts to explore, and the galaxies themselves reporting low SNR. This work develops the \zlbg pipeline, an alternative classification and redshift identification approach for faint LBG spectra to the current state of the art method \lbgnet in DESI.
We test the developed pipeline on representative samples of the DESI Run 2 pilot programs of LBG selections, and find that classification can be improved upon with a contrastive learning architecture, with respect to the original analysis. In particular, classification results improve AUC from $\sim0.988$ to $\sim0.997$. Redshift measurements are kept very much comparable to the performance displayed by \lbgnet. On the architecture side, the network infrastructure is kept as close as possible to \qnet, with the same backbone encoder. A projection head and contrastive loss are added, replacing the "line finder" of \lbgnet and \qnet. The contrastive loss uses continuous weighting to smooth out embeddings, and appendices \ref{sec:appendix:class_weighting} and \ref{sec:appendix:redshift_weighting} show adding relationship weights offers non-negligible improvements in classification and redshift performance. Appendix~\ref{sec:appendix:lz} explores the reported performance after using these preliminary redshift predictions in combination with \redrock. Finally, Appendix~\ref{sec:appendix:bad_spectra} describes the response of the pipeline to "BAD" spectra that do not give a robust redshift measurement. \zlbg also gives new information about the observed spectra, by returning confidence thresholds on the individual type of each galaxy, as showcased in Section~\ref{sec:results:classification}. Section~\ref{sec:results:joint} demonstrates that the confidence correlation between types is also fairly low, thus showing \zlbg leverages different information than \lbgnet. Therefore, a combination of both neural architectures to obtain further refinement of classification and redshift measurement performance would prove particularly useful in future works.

\paragraph{Products} The \zlbg pipeline is made available at \href{https://github.com/JeanCHDJdev/desi-lbg}{\textcolor{blue}{github.com/JeanCHDJdev/desi-lbg}}. Part of the DESI data used in this analysis will be made public alongside Data Release 2 (DR2) and DR3. DR1 and further release information can be found at: \href{https://data.desi.lbl.gov/doc/releases/}{\textcolor{blue}{data.desi.lbl.gov/doc/releases/}}. Data to recreate the figures of this work as well as a description of the provenance of the spectra used are available in \href{https://doi.org/10.5281/zenodo.21775480}{\textcolor{blue}{\texttt{zenodo}}}.
\acknowledgments
\label{sec:acknowledgements}

This material is based upon work supported by the U.S. Department of Energy (DOE), Office of Science, Office of High-Energy Physics, under Contract No. DE–AC02–05CH11231, and by the National Energy Research Scientific Computing Center, a DOE Office of Science User Facility under the same contract. Additional support for DESI was provided by the U.S. National Science Foundation (NSF), Division of Astronomical Sciences under Contract No. AST-0950945 to the NSF’s National Optical-Infrared Astronomy Research Laboratory; the Science and Technology Facilities Council of the United Kingdom; the Gordon and Betty Moore Foundation; the Heising-Simons Foundation; the French Alternative Energies and Atomic Energy Commission (CEA); the National Council of Humanities, Science and Technology of Mexico (CONAHCYT); the Ministry of Science, Innovation and Universities of Spain (MICIU/AEI/10.13039/501100011033), and by the DESI Member Institutions: \url{https://www.desi.lbl.gov/collaborating-institutions}. Any opinions, findings, and conclusions or recommendations expressed in this material are those of the author(s) and do not necessarily reflect the views of the U. S. National Science Foundation, the U. S. Department of Energy, or any of the listed funding agencies.

The authors are honored to be permitted to conduct scientific research on I'oligam Du'ag (Kitt Peak), a mountain with particular significance to the Tohono O’odham Nation.

This analysis used the following software packages: \texttt{numpy} \cite{harris2020numpy}, \texttt{SciPy} \cite{2020SciPyNMeth}, \texttt{matplotlib} \cite{Hunter2007Matplotlib}, \texttt{scikit-learn} \cite{ScikitLearnPedregosa2011}, \texttt{astropy}\cite{astropy2013paper1, astropy2018paper2, astropy2023paper3}, \texttt{pandas}\cite{mckinney-proc-2010-pandas, reback2020pandas}, \texttt{umap} \cite{mcinnes2020umapuniformmanifoldapproximation}, \texttt{TensorFlow} \cite{tensorflow2015-whitepaper} and \texttt{keras} \cite{chollet2015keras}.

\newpage
\appendix
\section{Redshift kernel weighting}\label{sec:appendix:redshift_weighting} 
\begin{figure}[h]
    \centering
    \includegraphics[width=0.98\linewidth]{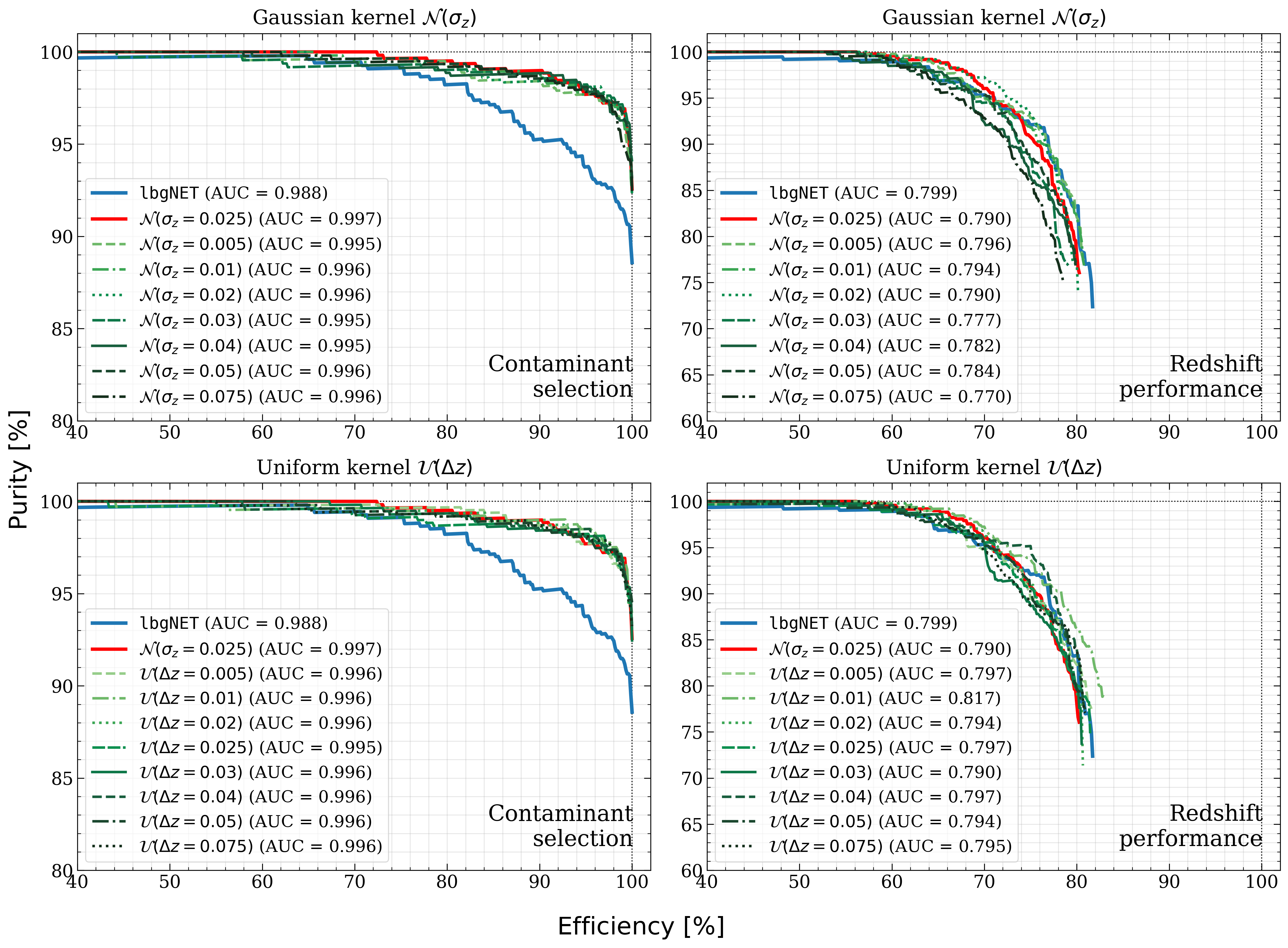}
    \caption{Contaminant selection and redshift performance purity/efficiency curves and their reported AUC for eight different Gaussian kernels ($\sigma_z=0.005, 0.01, 0.02, 0.025, 0.03, 0.04, 0.05, 0.075$ top figures), and top-hat kernels ($\Delta z$ using the same values, bottom figures). These are compared to the \lbgnet default performance (blue) using the "Base" training configuration, and the fiducial network for \zlbg which uses $\sigma_z=0.025$ (red).}
    \label{fig:appendix_weighting_z}
\end{figure}

In this appendix, we discuss some of the different redshift kernel training choices and their impact on the results. In particular, Figure~\ref{fig:appendix_weighting_z} showcases models using the same shared parameters, only changing the kernel (either top-hat $\mathcal{U}$ or gaussian $\mathcal{N}$). 

Figure~\ref{fig:appendix_weighting_z} shows the performance (with all the other parameters remaining fixed) of redshift kernels in the training weighting for the "Base" configuration. We observe results vary with no significant observed trend, although it seems like uniform kernels tend to return slightly better redshift performance overall. Such variance is possible when taking into account the small size of the training and test sets, and further refinements at the survey validation level would greatly help understand the appropriate kernel size. Nonetheless, based on the amplitude pre-factor of $\lz$ and the associated classification performance, the fiducial model retains $\sigma_z=0.025$ for now. We note that the performance is also dependent on the $\lz$ choice. In appendix~\ref{sec:appendix:lz}, we discuss the impact of $\lz$.

\section{Intra-class weighting}\label{sec:appendix:class_weighting}

\begin{figure}[h]
    \centering
    \includegraphics[width=0.98\linewidth]{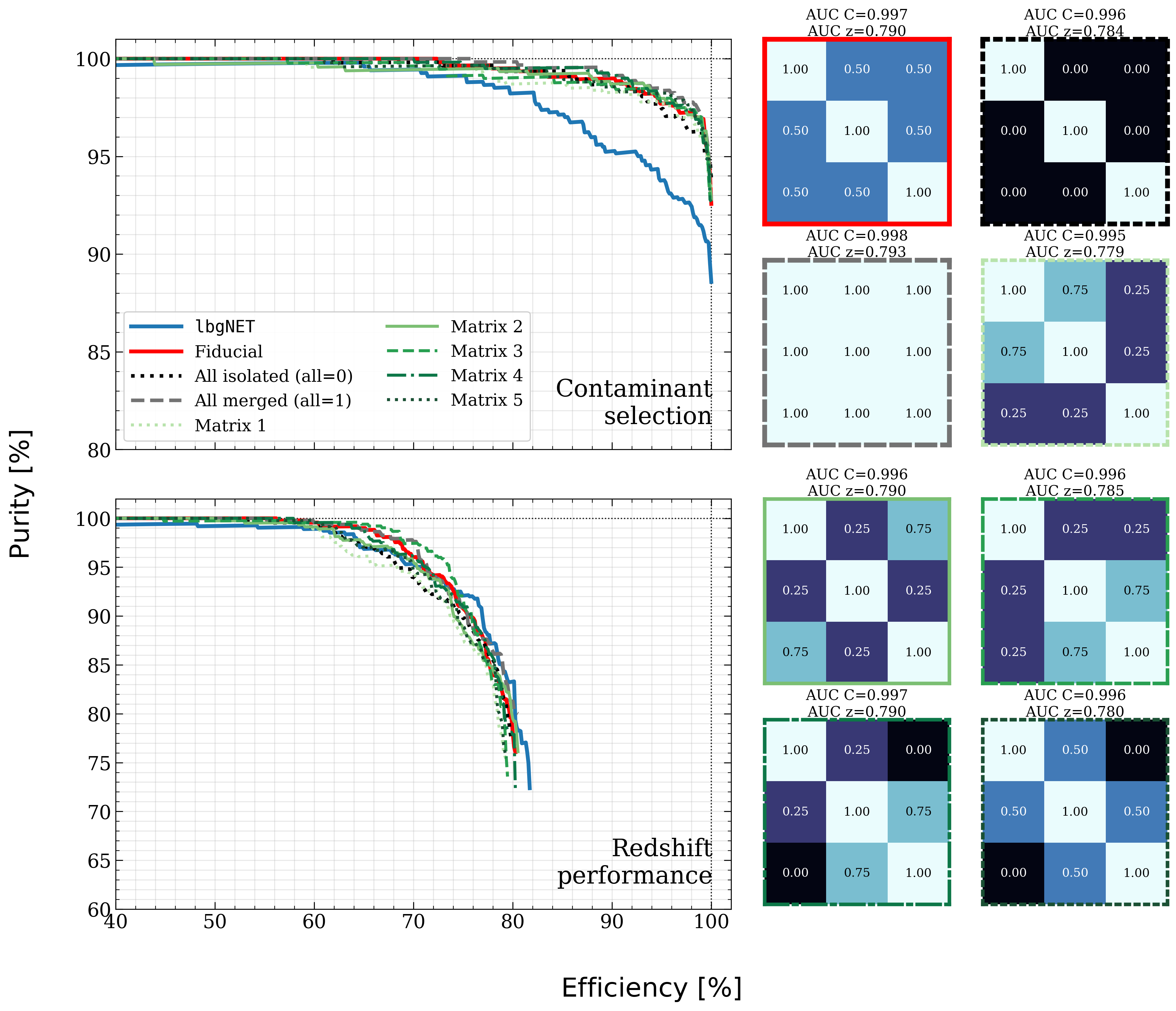}
    \caption{Contaminant selection (top left) and redshift performance (bottom left) purity/efficiency curves and their reported AUC for different $\mathbf{C}$-weight configurations. We restrict the matrix presentations to the LBG-type coefficients, since contaminant coefficients are entirely diagonal. By row (top to bottom) or column (left to right) the presented order is the following: LBGa, LBGme, LBGse. The particular cases of a diagonal matrix (\textit{All isolated}, black) and a all-ones matrix (\textit{All merged}, gray) are presented as well. These are compared to the \lbgnet default performance (blue) using the "Base" training configuration, and the fiducial network for \zlbg (red).}
    \label{fig:appendix_weighting_class}
\end{figure}

Another tunable parameter are intra-class weights, using the class relationships matrix defined in Section~\ref{sec:architecture:weights}. In this appendix, we investigate variations to the $c_{ij}$ coefficients. Figure~\ref{fig:appendix_weighting_class} presents some weighting configurations and their associated performance. Because we are prone to fine-tuning based on results on the test dataset, we pick a simple matrix, as defined in Equation~\ref{eq:fid_rel_matrix} (red in the figure). The matrices displayed focus on the intra-LBG weights (in order LBGa, LBGme, LBGse). ELGs and QSOs share no off-diagonal connection weights, therefore they are not shown in matrix coefficients.

Some matrix configurations present very good results, but ultimately these matrices are likely to depend on the structure of the test dataset. Overall, the takeaway should be that no intra-class weighting (black) performs worse in terms of redshift performance and classification than any simple weighting scheme, which motivates the $\mathbf{C}$ matrix. To a certain extent, this is also verified in the gray ("all merged") scenario, where all LBGs are considered to be perfectly the same: in this particular choice of $\lz$ and other hyperparameters, the all-merged configuration even nets the best results. Given the variance of the test dataset and the fluctuating results based on training seeds, we conserve our matrix of choice, since this still allows for classification of LBG sub types while retaining better performance than no LBG weights. Future improvements could include physically motivated proximity relationships; empirically, the well performing matrices show some intuitive physical connections. Taking the example of the bottom left matrix which returns good performance (AUC contaminants: 0.997, AUC redshifts: 0.790), one has $c_\mathrm{LBGse-LBGme}>c_\mathrm{LBGa-LBGme}>c_\mathrm{LBGa-LBGse}>c_\mathrm{\{LBG\}-\{ELG,QSO\}}$ where $c_{ij}$ are the connection coefficients. One expects the strong and moderate emission to be structurally close, whereas absorption and strong emission are expected to be further apart. These observations can only really be confirmed with larger testing and training sets; hence this is left for future works.

\section{Redshift recovery at varying prior sizes}\label{sec:appendix:lz}

This appendix inspects the redshift recovery performance after applying the $\lz$ priors to the dataset. At the fiducial $\lz$, \lbgnet and \zlbg show comparable performance. However, \lbgnet is inherently built to detect spectroscopic lines. Intuitively, by construction, \lbgnet is more prone to catastrophic outliers (for example, by misidentifying lines). On the contrary, the \zlbg pipeline aligns the embeddings by redshift proximity, not specifically looking for emission lines in $\textbf{Rep}$ space. This means redshifts should on average closer to their true value with \zlbg than with \lbgnet. The behavior is well characterized by tuning the failure point $\lz$ for redshift, allowing more leniency for catastrophic outliers.

\begin{figure}[h]
    \centering
    \includegraphics[width=0.98\textwidth]{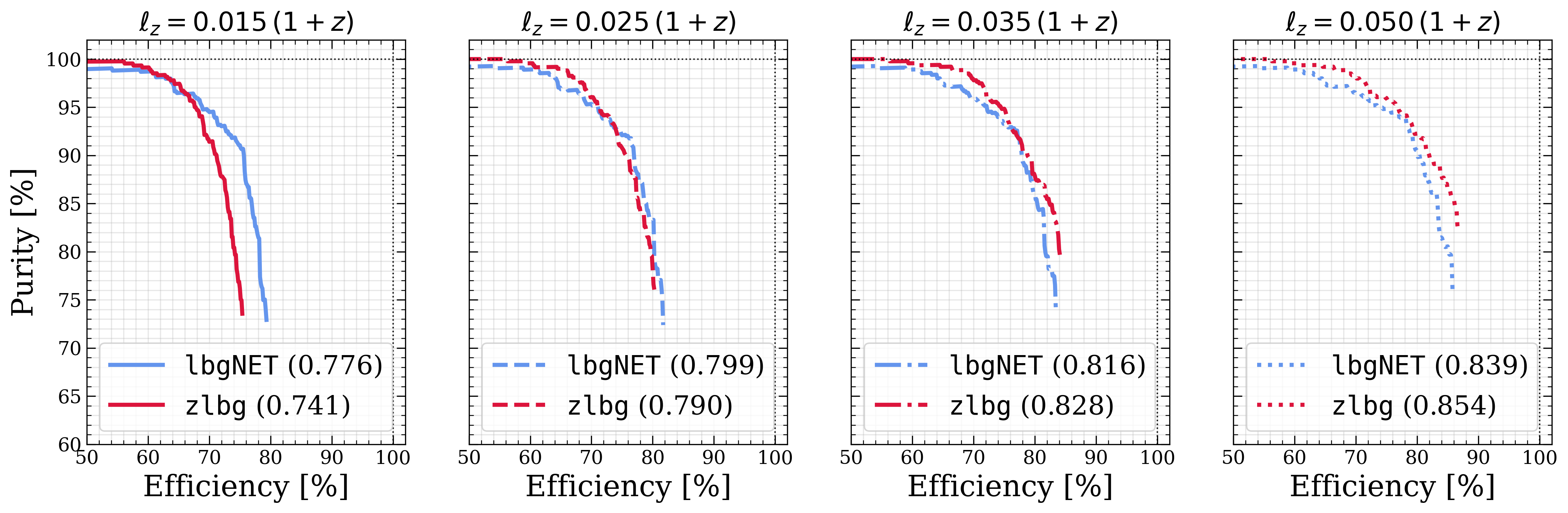}
    \caption{Redshift performance as $\lz$ (prior half-width for \redrock) varies. The networks used are the same for every plot (fiducial configuration, with Base training set).}
    \label{fig:results:lz_variation}
\end{figure}

Figure~\ref{fig:results:lz_variation}, shows that increasing $\lz$ further increases the redshift performance of \zlbg compared to \lbgnet, reaching better AUC at and above $\lz=0.035(1+z)$. The failure point evolves for both networks, but the performance gain on \zlbg is larger. The classification performance remains independent of $\lz$, so the only limiting factor to changing the prior half-width is whether the recovered redshift after \redrock is still correct despite a larger prior exploration range.

\begin{figure}[h]
    \centering
    \includegraphics[width=0.85\textwidth]{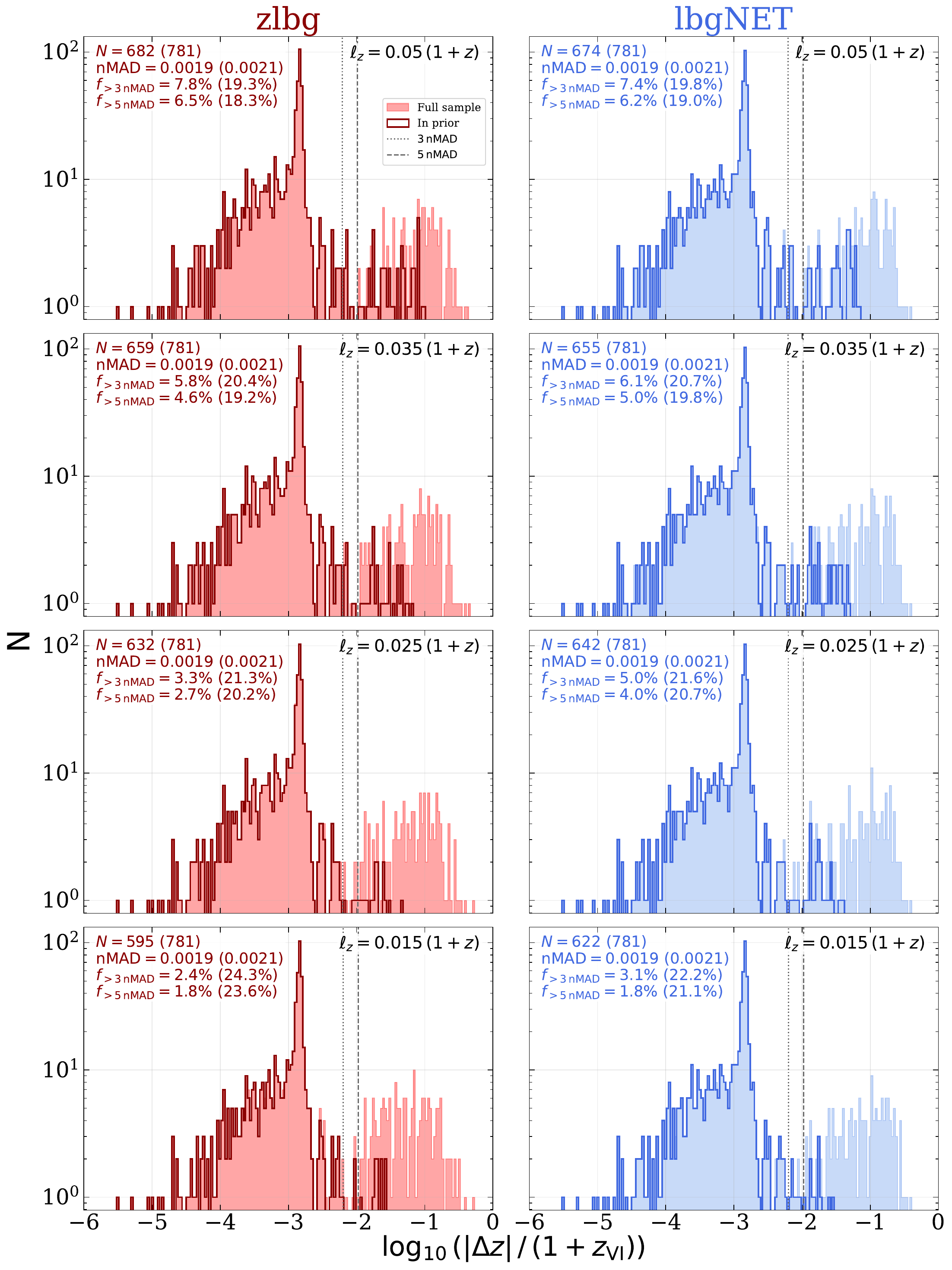}
    \caption{\redrock residuals compared to the visual inspection measurements for \zlbg (top) and \lbgnet (bottom) using $\log(|z_\mathrm{VI}-z_\mathrm{rr}|/(1+z_\mathrm{VI}))$. We show the normalized median absolute deviation (nMAD), outlier fraction at 3nMAD ($f_\mathrm{>3nMAD}$) and at 5nMAD ($f_\mathrm{>5nMAD}$), and the number of recovered sources. In parentheses, we show the results for the full sample. The outlined histogram corresponds to the sources that fall within the prior, that is $|z_\mathrm{VI}-z_\mathrm{rr}|<\lz$.}
    \label{fig:appendix:redrock}
\end{figure}

This is explored in Figure~\ref{fig:appendix:redrock}.
\redrock is ran on the co-added spectra of the test dataset. The templates used for \redrock are the ones described in Section~\ref{sec:data:templates}. We report the log-residual values of the visually inspected redshift with the redshift measured by \redrock ($\log(|z_\mathrm{VI}-z_\mathrm{rr}|/(1+z_\mathrm{VI}))$), both for the full sample of test LBGs (781 galaxies, filled histogram) and the sample that fall into the prior range (outlined histogram). The bi-modal distribution shows the redshift success (within $\sim10^-2$ of the VI value). While such tests are subject to the structure of the test dataset, it is interesting to note that there is in general a lower contamination rate for \zlbg. nMAD remains quite similar over every test. The contamination rate also increases as a function of the amplitude pre-factor of $\lz$, as expected: however, the number of sources recovered also increases significantly.


\begin{figure}[h]
    \centering
    \includegraphics[width=0.98\textwidth]{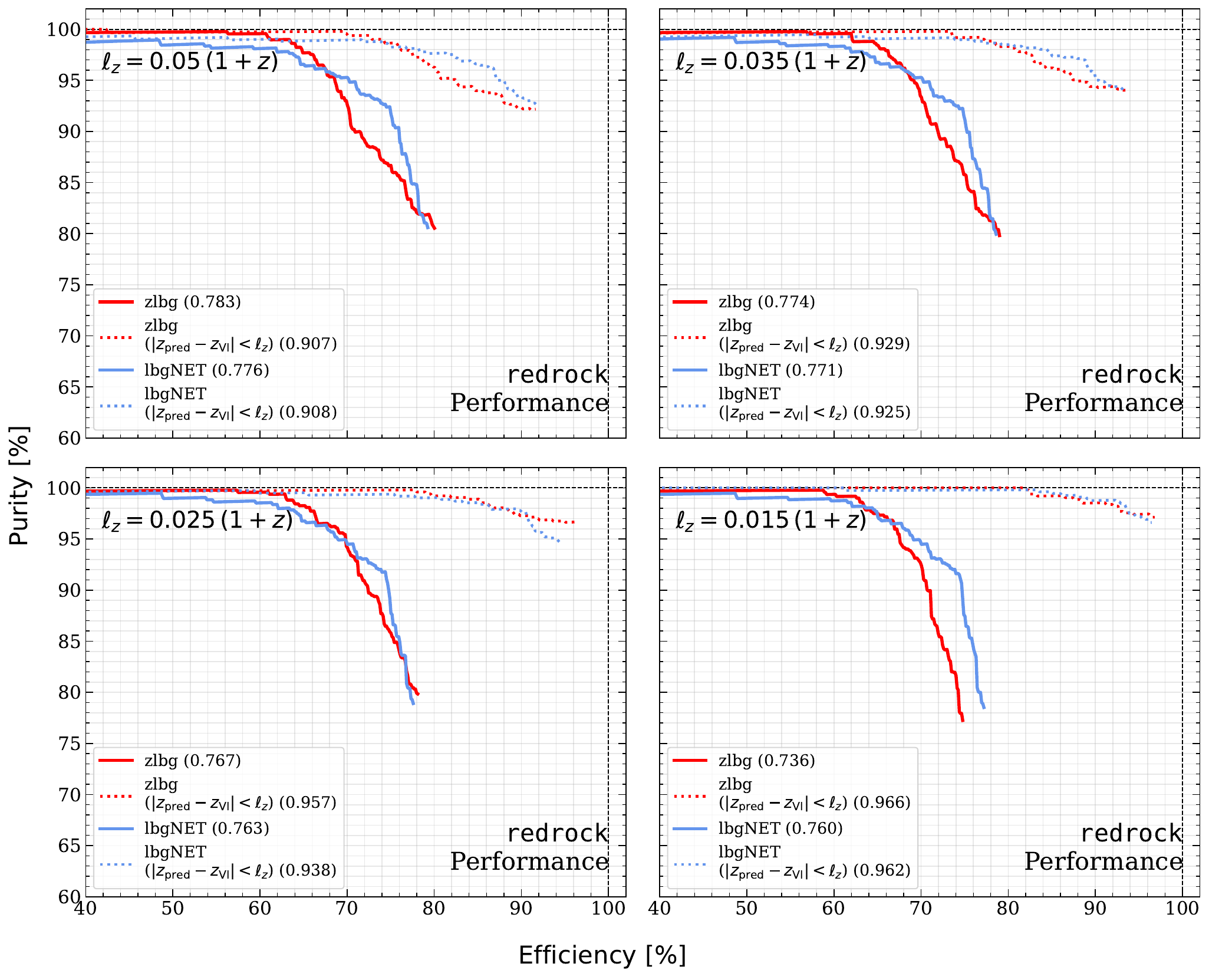}
    \caption{Efficiency/purity curves for \redrock efficiency. In dotted, we show the \redrock performance over only the sources that fall within the $\lz$ prior. The efficiency/purity curve for the end-to-end inference (restricted to the LBGs) is reported in full line: the failure threshold is $|z_\mathrm{VI}-z_\mathrm{rr}|>\epsilon_\mathrm{rr}(1+z_\mathrm{VI})$ after the prior prediction. Here, $\epsilon_\mathrm{rr}=0.005$ is the tolerated offset with $z_\mathrm{VI}$. The dotted line showcases specifically the failures for sources inside the prior. The AUC is reported in parentheses.}
    \label{fig:appendix:redrock_effpur}
\end{figure}

In Figure~\ref{fig:appendix:redrock_effpur}, we display the end-to-end (full line) and "in prior" (dotted) efficiency/purity curves over redshift. Broader priors show more outliers for sources that are in the prior before running \redrock, as one expects, but these larger priors also recover more sources, as shown in Figure~\ref{fig:appendix:redrock}. Another trend that can be observed, as hinted at by Figure~\ref{fig:results:lz_variation}, is the improvement of \zlbg at larger $\lz$ compared to \lbgnet, where AUC is more stable than \zlbg. For a strict $\lz=0.015(1+z)$ prior, the \zlbg AUC is much lower, but catches up with \lbgnet performance as the prior broadens. A compromise value for the prior amplitude is left for future work, particularly when a larger test and training set is assembled for meaningful statistics.

\section{Treatment of bad spectra}\label{sec:appendix:bad_spectra}

\begin{figure}[h]
    \centering
    \includegraphics[width=0.98\linewidth]{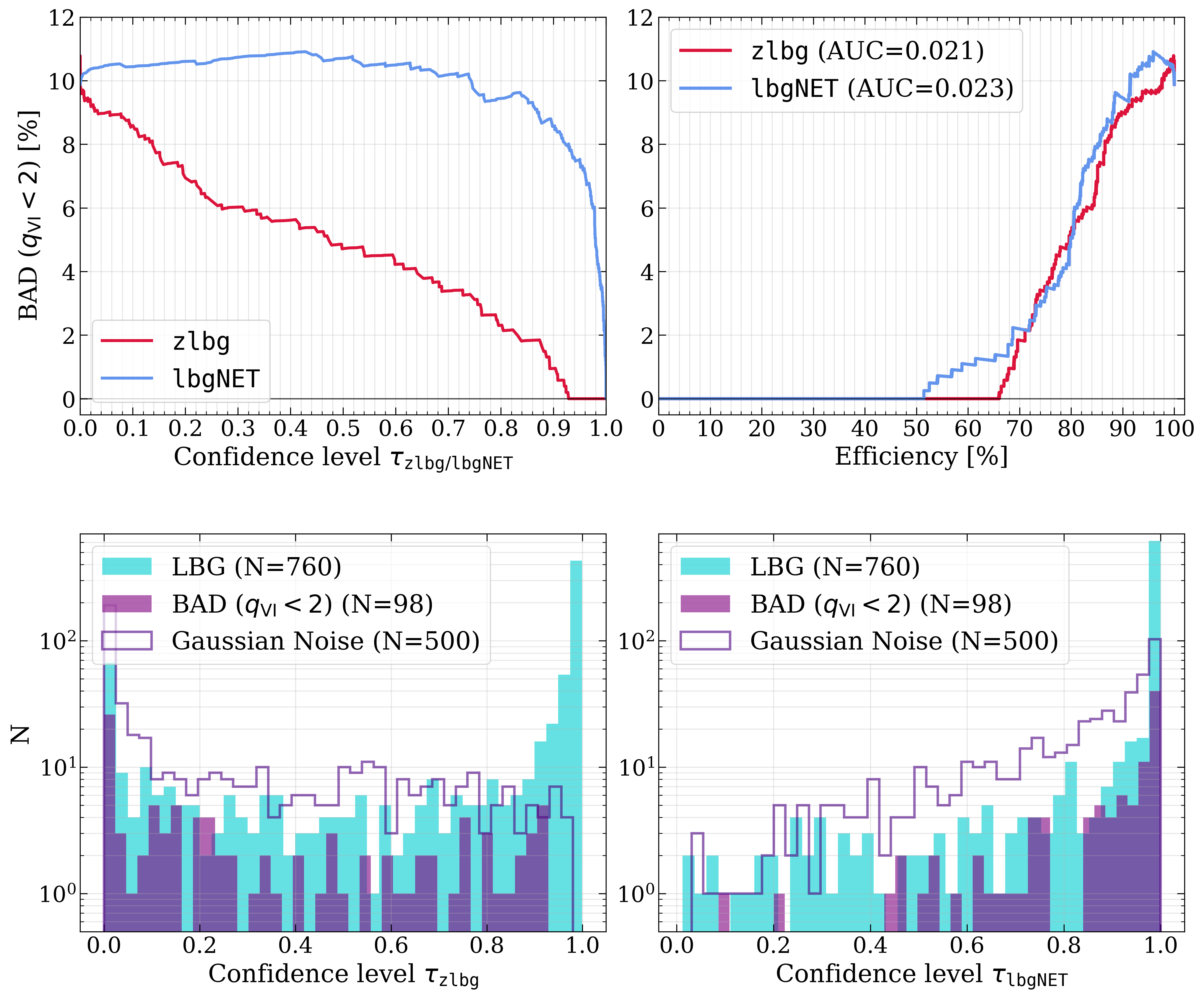}
    \caption{\textit{Top left}: Percentage of "BAD" spectra as a function of confidence level for both pipelines. \textit{Top right}: Percentage of "BAD" spectra as a function of LBG efficiency. \textit{Bottom}: Distribution of confidence levels for both \zlbg and \lbgnet for the "BAD" spectra, the "gaussian noise" mock spectra, and the LBGs.}
    \label{fig:bad_lo2}
\end{figure}

The LBG population observed is made of very faint galaxies with low SNR. For many reasons (not enough observation time, faint target, detector issue...) a spectra does not always give a good redshift. Due to this, visual inspection campaigns have flagged spectra considered as "BAD"; that is, when the assertion of redshift is not confident enough or the spectra is largely noise dominated compared to the noise of the detector. Such spectra are usually reported in VI by a quality flag $q_{\mathrm{VI}}$. 
Following \cite{VIGalaxies.Lan.2023, VIQSO.Alexander.2023, RuhlmannKleider2024_LBGs}, the VI quality flag describes: $q_{\mathrm{VI}}=3,4$ for secure redshifts, $q_{\mathrm{VI}}=2$ for low confidence redshifts and $q_{\mathrm{VI}}<2$ for no redshift assignment.  
$q_{\mathrm{VI}}$ is the average of the $q_{\mathrm{VI}}$ reported by the three reviewers if answers are compatible, or in certain case the flag attributed upon re-inspection to solve inconsistencies \cite{RuhlmannKleider2024_LBGs}. 

In this work, we select spectra with $q_{\mathrm{VI}}>2.5$. On the contrary, this appendix specifically focuses on the spectra considered bad or noisy, and thus we select all the spectra of our test dataset that report $q_{\mathrm{VI}} <2$: no redshift reported. These spectra are often noise or weak continuum with no discernible features. Therefore, we focus our analysis on the spectra with low enough quality that they should not fall into the selection, since $q_{\mathrm{VI}}\gtrsim2$ can return spectra with discernible redshift measurements. This selects 98 "BAD" spectra to test on. Moreover, we also draw 500 "gaussian noise" mock spectra, which are pure noise realizations based on the median inverse variance of these 98 bad spectra. Figure~\ref{fig:bad_lo2} showcases the contamination rate as a function of confidence level and LBG efficiency (top) and the distribution of confidence levels as a function of the datasets. Contamination remains acceptable even at high efficiencies, thus the network will likely correctly eliminate the spectra that are considered "BAD" for redshift measurements. In this scenario, contamination rate is better if the reported AUC in the top left plot is better.

\bibliographystyle{JHEP}
\bibliography{src/lbgs, src/methods, src/desi, src/software, src/surveys}

\end{document}